\documentclass[acmtog,authorversion]{acmartarxiv}
\usepackage{booktabs} 

\setcitestyle{square}

\newcommand{\nonarXiv}[1]{#1}
\renewcommand{\nonarXiv}[1]{}

\newcommand{\arXiv}[1]{#1}

\usepackage[ruled]{algorithm2e} 

\SetAlFnt{\small}
\SetAlCapFnt{\small}
\SetAlCapNameFnt{\small}
\SetAlCapHSkip{0pt}
\IncMargin{-\parindent}

\acmYear{2017}
\acmMonth{5}

\nonarXiv{\setcopyright{acmcopyright}}
\arXiv{\setcopyright{rightsretained}}

\usepackage[utf8]{inputenc}
\usepackage{amsmath}
\usepackage{float}
\usepackage{bm}
\usepackage{amsfonts}
\usepackage{color}
\usepackage{tabularx}
\usepackage{mathtools}
\usepackage{multirow}
\usepackage{siunitx}

\definecolor{lightgray}{rgb}{0.8,0.8,0.8}
\definecolor{darkblue}{rgb}{0,0,0.5}
\definecolor{darkyellow}{rgb}{0.4,0.3,0}
\definecolor{darkgreen}{rgb}{0,0.4,0}
\definecolor{darkpurple}{rgb}{0.5,0.0,0.4}
\definecolor{midred}{rgb}{0.8,0,0}
\definecolor{darkred}{rgb}{0.4,0,0}
\definecolor{midgreen}{rgb}{0,0.8,0}
\definecolor{midorange}{rgb}{0.8,0.6,0}
\definecolor{turquoise}{rgb}{0,0.6,0.8}

\newcommand{\uvec}[1]{{\hat{\bm{#1}}}}
\newcommand{\cvec}[1]{{\bm{#1}}}

\newcommand{\Sec}[0]{Sec.}
\newcommand{\App}[0]{App.}
\newcommand{\Eq}[0]{Eq.}
\newcommand{\Fig}[0]{Fig.}
\newcommand{\scratchpos}{{\cvec{r}}}

\newcommand{\incident}{i}
\newcommand{\tancoord}{t}
\newcommand{\scratch}{\mathrm{scratch}}
\newcommand{\base}{\mathrm{base}}
\newcommand{\mask}{\mathrm{mask}}

\newcommand{\hide}[1]{}

\title{Scratch iridescence: Wave-optical rendering of diffractive surface structure}
\author{S. Werner}
\affiliation{Rheinische Friedrich-Wilhelms-Universit{\"a}t Bonn, Germany}
\author{Z. Velinov}
\affiliation{Rheinische Friedrich-Wilhelms-Universit{\"a}t Bonn, Germany}
\author{W. Jakob}
\affiliation{Ecole Polytechnique F\'ed\'erale de Lausanne (EPFL), Switzerland}
\author{M.\,B. Hullin}
\affiliation{Rheinische Friedrich-Wilhelms-Universit{\"a}t Bonn, Germany}

\begin{document}
\begin{teaserfigure}
 \centering
  \includegraphics[width = 0.33\linewidth]{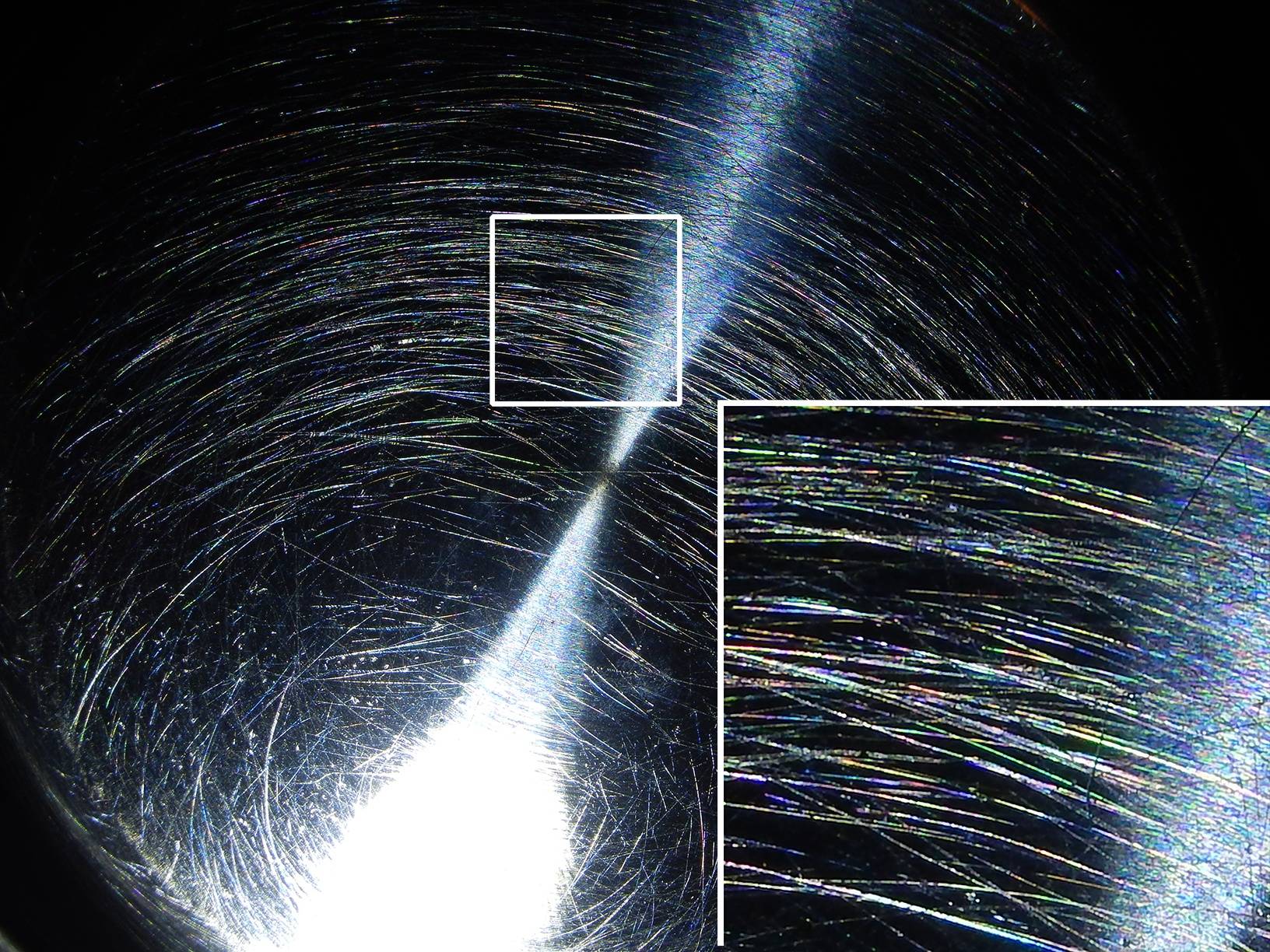}\hfill
  \includegraphics[width = 0.33\linewidth]{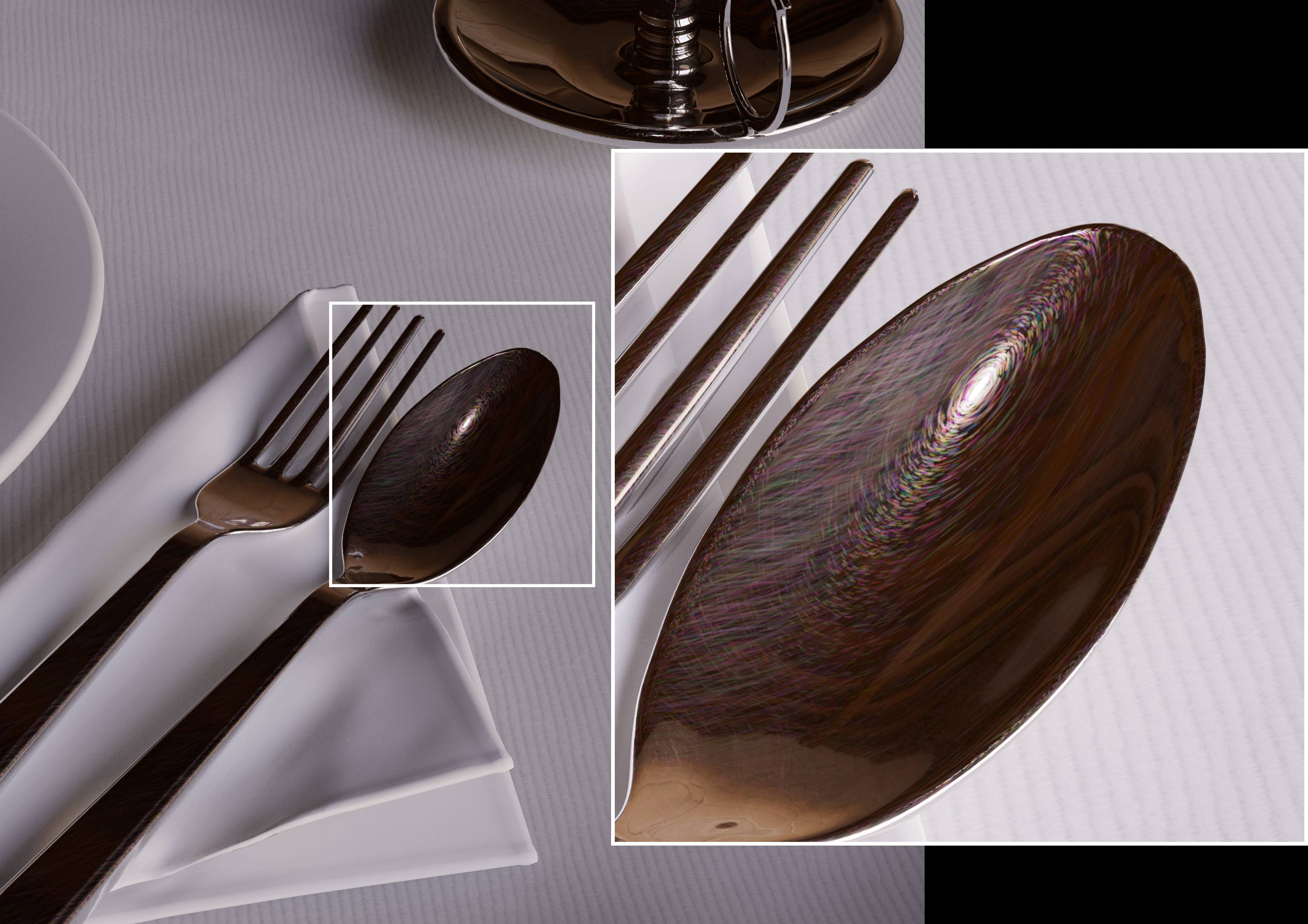}\hfill
  \includegraphics[width = 0.33\linewidth]{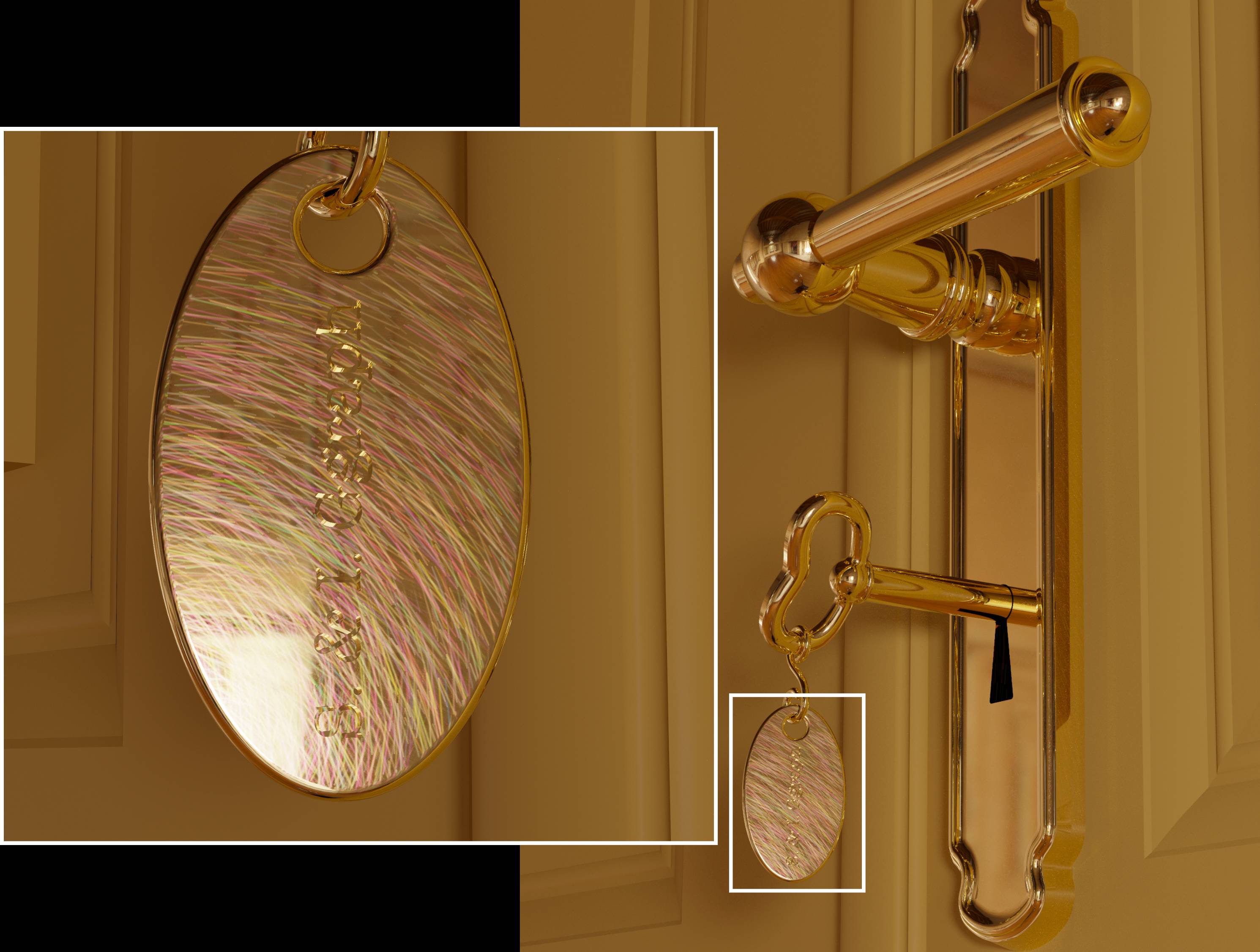}
   \caption{Left: Inside of a cooking pot photographed under a halogen spotlight. Center and right: Renderings of scratched materials under environment lighting obtained using our model. }
 \label{fig:teaser}
\end{teaserfigure}

\begin{abstract}
The surface of metal, glass and plastic objects is often characterized by microscopic scratches caused by manufacturing and/or wear. 
A closer look onto such scratches reveals iridescent colors with a complex dependency on viewing and lighting conditions. 
The physics behind this phenomenon is well understood; it is caused by diffraction of the incident light by surface features on the order of the optical wavelength. 
Existing analytic models are able to reproduce spatially unresolved microstructure such as the iridescent appearance of compact disks and similar materials. 
Spatially resolved scratches, on the other hand, have proven elusive due to the highly complex wave-optical light transport simulations needed to account for their appearance. 
In this paper, we propose a wave-optical shading model based on non-paraxial scalar diffraction theory to render this class of effects. 
Our model expresses surface roughness as a collection of line segments. To shade a point on the surface, the individual 
diffraction patterns for contributing scratch segments are computed analytically and superimposed coherently. 
This provides natural transitions from localized glint-like iridescence to smooth BRDFs representing the superposition of many reflections at large viewing distances. 
We demonstrate that our model is capable of recreating the overall appearance as well as characteristic detail effects observed on real-world examples.
\end{abstract}

\maketitle

\section{Introduction}\label{sec:introduction}
Modelling and rendering the real world with all its irregularities and imperfections remains one of the greatest challenges in computer graphics. A rich history of research on dirt, dust and fingerprints, weathering, patination, erosion \cite{dorsey2010digital}
and scratches \cite{Merillou2001,Bosch2004,Dong2015,Yan2016,Raymond2016} documents the massive amount of effort invested by our community to make computer graphics look less sterile and more realistic.
In this paper, we focus on a subtle but very common effect observed on objects made of metal, glass or plastic. Under strongly directional lighting (like sunlight or a halogen spot), these surfaces exhibit colorful patterns that are caused by diffraction of light reflecting
off microscopic surface details (\Fig~\ref{fig:teaser}). Being fundamentally a wave-optical phenomenon, this effect cannot be reproduced
by geometric optics models and requires a careful study of both diffraction by
individual surface features at the microscopic scale as well as interference among
multiple features. The most detailed wave-optical simulations conducted in computer graphics thus
far have involved detailed finite-difference time-domain (FDTD) modeling of
periodic microstructures on butterfly wings \cite{Musbach2012}, an approach that is not feasible for structures beyond a few cubic micrometers.\footnote{For instance, FDTD
simulation of light waves in a 1\,mm$^3$ volume would require
2.4$\times$10$^{28}$ cell updates  per second of simulated time ($\lambda=500$\,nm, grid resolution
$\lambda/10$, time resolution $T/10$).} In other works, the appearance of large-scale diffractive objects has been approximated by combining far-field scattering models for repetitive microscopic structures with traditional texturing approaches.
Most objects with which we interact in our everyday use, however, exhibit features across many scales, ranging from macroscopic ones that are resolvable with the naked eye to microscopic ones that are only indirectly visible due to their aggregate interaction with light. 
This complexity leads to an intricate variation of appearance along the spatial, angular and spectral dimensions that no model so far has been able to express. Our model contributes a first step to 
provide a framework for rendering such phenomena.

To be able to simulate the appearance of scratches within modern physically based rendering systems, we propose a new bidirectional reflectance distribution function (BRDF) modeling surfaces covered with microscopic scratch particles.
This involves a number of challenges: for instance, traditional analyses of the diffracted wave field rely on a paraxial (small-angle) assumption that would lead to grossly inaccurate results in a BRDF model that must support evaluation for any pair of incident and outgoing angles.
Our model thus simulates the diffraction of light by microscale surface features using a \emph{non-paraxial} scalar diffraction theory proposed by Harvey~\shortcite{Harvey2000}.
Our formulation leads to a BRDF with naturally coupled spatial, angular and spectral variation that 
exhibits multi-scale behaviour: at a distance, interference from a larger number of scratches causes it to resemble standard geometric optics models, while larger magnifications reveal iridescent reflections from small collections of scratches.

To model and simulate rough and scratched surfaces, we propose a vector graphics representation where an ensemble of fundamental primitives (linear scratch segments) resides on a base substrate. 
We show how the constituent scattering distributions can be expressed in closed form, and discuss automated and interactive techniques for placing large numbers of scratches on surfaces.
Finally, we integrate our model into a modern physically-based rendering system and discuss efficient implementations of key operations, including importance sampling.

%
\section{Related work}\label{sec:related-work}
Detailed modeling and rendering of surface defects can dramatically improve the
realism of renderings, hence the pursuit of such models has been a topic of
great interest to the rendering community at large. Early work in this area
includes methods by Buchanan and Lalonde~\shortcite{Buchanan1999} and Lu et
al.~\shortcite{Lu2000} who analyze general reflection properties of scratches.
More recent work on rendering scratched surfaces can generally be grouped into
three high-level categories.

\textbf{Explicit geometry.} Merillou et al.~\shortcite{Merillou2001} and Bosch
et al.~\shortcite{Bosch2004,Bosch2008} consider the geometry of scratches to
derive BRDFs taking into account the scratch cross-section (profile) at each
shading point. Merillou et al.~use a preset profile and a texture mapping to
position scratches, with scratch profiles split up into a number of
symmetrically arranged tilted surfaces with associated procedural BRDFs. Bosch
et al.~enhance this model with generalized profile representations and curves
on the surface of shaded objects to position scratches.
Raymond et al.~\shortcite{Raymond2016} propose a multi-scale SVBRDF model
based on mirror-like scratches organized in a set of coherently oriented
scratch layers; their model relies on an accurate solution of interreflection
within a scratch and supports multi-scale evaluation. All of these approaches
separate spatial and optical information concerning the scratches, and their
solutions are only valid in the geometrical optics regime.

\textbf{Microfacet models.} BRDF models based on microfacet theory are widely
used in graphics and have proven effective in reproducing the appearance of
real-world materials ~\cite{Ngan2005b}. The key quantity of a microfacet BRDF
is its normal distribution function (NDF), which provides a direct link between
the model's scattering behavior and the statistics of an underlying micro-scale surface structure. Many
NDF models have been proposed in the past, hence we only focus on ones that
specifically target rendering of scratched surfaces. Yan et al.~\shortcite{Yan2014}
numerically integrate the NDF of normal-mapped surfaces over the surface region
observed within a single pixel, which yields an efficient multi-scale
reflectance model capable of rendering high-resolution normal maps under
directionally peaked illumination. A later work~\cite{Yan2016} further enhances
performance using four-dimensional position-normal distributions represented as
Gaussian mixtures.

Dong et al.~\shortcite{Dong2015} compare the use of microfacet and Kirchhoff
scattering theory to predict surface roughness from measured microgeometry.
Both theories yield highly accurate reflectance models, but the use of a
profilometer limits their technique to BRDFs as only small surface patches can
be measured. The authors show that SVBRDFs of brushed aluminium can be
synthesized using a texture-based approach but do not investigate spatially
resolved scratches.

\textbf{Diffraction.}
{A limitation that all existing models and techniques for scratches share is their restriction
to geometric optics. They therefore cannot account for wave-optics phenomena such as 
diffraction and interference of nearby microscopic scratches or, in general, surface roughness.
Modelling diffraction by rough surfaces has been of great interest to the physics community and 
various models addressing the different characteristics of surfaces have been developed, ranging from Rayleigh-Rice vector perturbation theory (smooth surfaces) to more general ones such as Beckmann-Kirchhoff scattering theory (variety of surface roughness classes but only small angles). A good overview of these scattering theories and the development of more recent ones can be found in Krywonos~\shortcite{Krywonos2006}.

In the computer graphics community, a variety of BRDF models have been developed to account for diffraction effects created by microscale surface
features in ray-based frameworks. However, none of them targets the transition between texture and far-field diffraction that is needed for our purpose. One of the first to incorporate
scattering theory based diffraction into a BRDF model were Church and
Takacs~\shortcite{Church1995}. L\"ow et al.~\shortcite{Loew2012} later
introduced their ABC model to the graphics community and demonstrated its
merits in numerical fits to measured reflectance data.
He et al.~\shortcite{He1991} derived a BRDF model based on vector Kirchhoff
theory for surfaces with roughness described by a Gaussian random process.
Stam~\shortcite{Stam1999} proposed a BRDF model based on scalar
Kirchhoff theory, capable of rendering the diffraction effects of random distributions of primitives (such as rectangular bumps) or 
Gaussian random surfaces by utilizing the power spectrum of the autocorrelation function of the surface height variations.

Sun et al.~\shortcite{Sun2000} derive an accurate far-field diffraction model
to render compact discs modeled as a series of concentric tracks with a quasi-periodic arrangement of pits. 
Ritschel et al.~\shortcite{Ritschel2009} as well as Hullin et al.~\shortcite{HullinSIG2011} reproduce 
wave-optical effects such as glare (scattering of light within the human eye) or lens flare.

Cuypers et al.~\shortcite{Cuypers2012} propose a Wave Bidirectional Scattering
Distribution Function (WBSDF) based on statistical optics,
computing the Wigner distribution function
(WDF) of microstructures to produce solutions that are valid in the near and far field.
The WBSDF approach relies on analytic solutions for simple quasi-periodic
structures and falls back to precomputed lookup tables for the general case of
arbitrary discretized microstructure. However, this precomputation does not
scale to complex non-periodic microstructures due to the exceedingly high
memory requirements associated with the underlying four-dimensional
representation. More recently, Dhillon et al.~\shortcite{Dhillon2014} developed a 
diffraction model based on heightfield data acquired using an atomic-force microscope (AFM);
this data was used to generate lookup-tables for interactive rendering by truncating a Taylor-series expression 
of the BRDF. To better reproduce surface scattering effects based on statistical properties of heightfields, Holzschuch and Pacanowski~\shortcite{Holzschuch2016}
introduced the generalized Harvey-Shack theory to the computer graphics community.

Musbach et al.~\shortcite{Musbach2012} proposed a reflectance model for
iridescent biological structures based on a detailed FDTD simulation of
vectorial wave propagation. This approach is significantly more general than
the previously discussed models, but the prohibitive cost of this type of
simulation limits it to quasi-periodic structures. Levin et al.~\shortcite{Levin2013} use scalar Kirchhoff theory to reproduce
BRDFs of a specific class of surfaces for photolithographic reconstruction.
While only remotely related to our work, their approach is one of the few to
consider the effects of spatial coherence of the illumination source.

\textbf{Image-based techniques.}
Also related to our work are image-based techniques that fit anisotropic
reflectance models to measurements
of finished wood~\cite{Marschner2005} or brushed metal surfaces~\cite{Wang2008,Yue2010}.

To overcome the discussed restrictions in the context of rendering scratched
surfaces, our approach builds on an efficient representation tailored to this
application. Similar to prior work \cite{Merillou2001,Bosch2004}, we separate
spatial and optical information by describing the scratch layout as a curve and
its reflectance behavior using a profile at each position along the curve. We
use Harvey's non-paraxial scalar diffraction theory~\shortcite{Harvey2000} to
express the diffracted reflectance as a superposition of reflections from
individual scratches. Similar to Sun et al.~\shortcite{Sun2000} we derive the
BRDF from the explicit calculation of the scattered complex wavefront,
maintaining as much generality as possible. This allows us to take into account
spatial coherence similar to Levin et al.~\shortcite{Levin2013} to reproduce
not only diffraction effects but also the mutual interference created by dense
scratch ensembles.
%
\begin{figure}[t]
\centering
\includegraphics[width=0.7\linewidth]{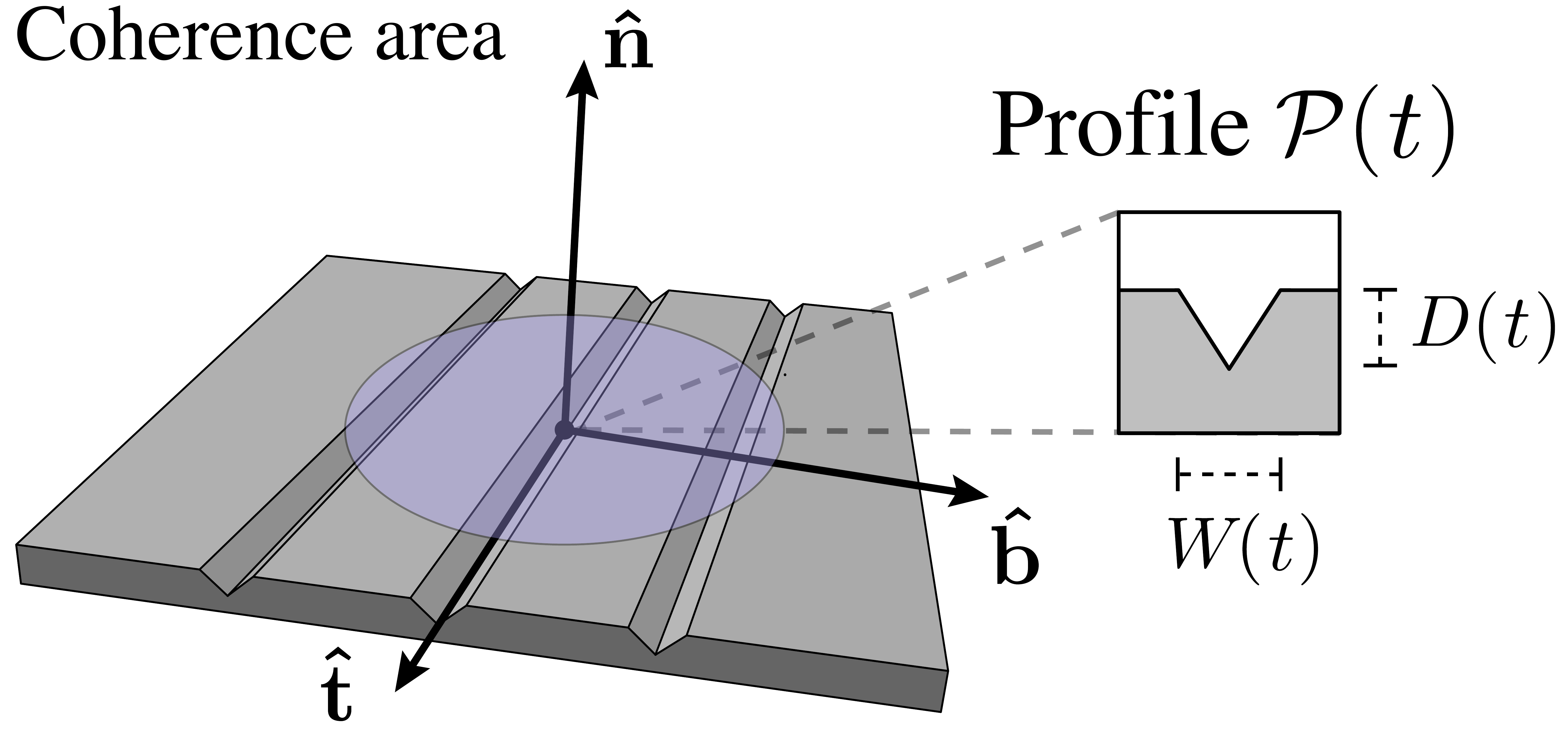}\hfill%
\caption{
Local shading geometry: Scratches that lie within the coherence area contribute to the diffracted radiance. We represent each scratch by a parametric curve $\scratchpos(\tancoord)$, which brings its own local coordinate system with tangent and bitangent directions $\uvec{t}(\tancoord)$ and $\uvec{b}(\tancoord)$, respectively. The cross-section at any position $\tancoord$ along the scratch is defined by a profile $\mathcal{P}(\tancoord,b)$. We use profiles with analytical Fourier transforms and scale them in the ($\uvec{b}$-$\uvec{n}$)-plane using parameters $W$ (width) and $D$ (depth)}%
\label{fig:shading-geometry}%
\end{figure}
\begin{figure*}[t]%
\centering
\includegraphics[width=.99\linewidth]{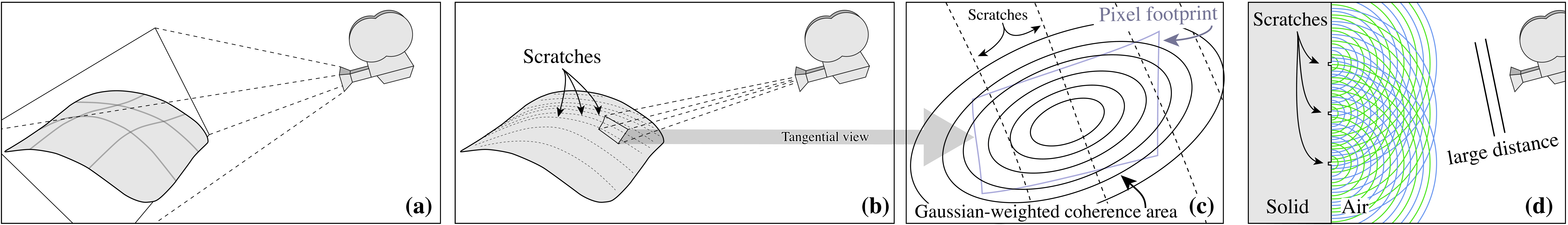}
\vspace{-2mm}
    \caption{From scratches to diffracted radiance:
    \textbf{(a)} Macroscopic view of a surface containing scratch particles.
    \textbf{(b)} Ray differentials establish a mapping between the visible surfaces and pixels of the output image.
    \textbf{(c)} Our method considers scratches located inside a Gaussian-weighted coherence region around the surface region covered by a pixel.
    \textbf{(d)} Side view: each scratch in the coherence region scatters the
    incident light into wavelets that interfere with each other; although not
    shown here, our method also accounts for reflection from the non-scratched
    base surface. Since the distance to the camera is much greater than the
    wavelength of light, it is enough to consider diffracted radiance, a
    far-field approximation of the superposition of wavelets that remains
    accurate for all angles of observation.}
\label{fig:pixelzoom}%
\end{figure*}
\begin{table}[t]
\centering
{\small\begin{tabular}[T]{|>{$}l<{$}|c|r|}
\hline
 \textbf{Symbol} & \bf Meaning & \bf{Reference}\\
 \hline
 {\cvec{x}}, {\uvec{x}} & Vector, unit vector &  \\
 \hline
 \uvec{\omega}_i, \uvec{\omega}_o & Incident and outgoing light directions &  \\ \hline\hline
 \multicolumn{3}{|c|}{\bf Spatial parameterization}\\\hline
\cvec x\!=\!(x,y,z)\!\!&Position in space&\\
 U(\cvec{x}) & Scalar field amplitude & \!\!Sec.~\ref{sec:preliminaries:diffrad}, Fig.~\ref{fig:angular-spectrum}a\\
 U_0(x,y) & Scalar field in plane $z=0$ & \\
 \hline\hline
 \multicolumn{3}{|c|}{\bf Plane-wave parameterization (spatial-frequency domain)}\\\hline
 (\alpha,\beta,\gamma) & Vector of direction cosines &\multirow{2}{*}{\Fig~\ref{fig:angular-spectrum}b, \Sec~\ref{sec:preliminaries:diffrad}}\\
 V(\alpha,\beta,z) & $z$-slice of field in spatial frequencies & \\
 \cvec{\xi} & $(\uvec{\omega}_o - \uvec{\omega}_i)/{\lambda}$ & Sec.~\ref{theory-spectral-brdf}\\
 \cvec{\xi}' & Projection of $\cvec{\xi}$ in scratch frame & \Eq~\ref{eq:scratch-frame-trafo}\\
 \hline\hline
 \multicolumn{3}{|c|}{{\bf Scratch representation}}\\\hline
 \tancoord & Position along the scratch & \multirow{6}{*}{\Sec~\ref{sec:preliminaries:scratch}, \Fig~\ref{fig:shading-geometry}}\\
 \scratchpos(\tancoord) & Point on scratch at position $\tancoord$ &  \\
 \multirow{2}{*}{$\{\uvec{t},\uvec{n},\uvec{b}\}(\tancoord)$} & Local scratch coordinate frame  & \\[-.8mm]
&(tangent, normal, bitangent) at $\tancoord$&\\
b&Bitangential coordinate&\\
 \mathcal{P}(\tancoord,b) & Scratch profile at $\tancoord$ as function of $b$ & \\
 W(\tancoord),D(\tancoord) & Scratch width and depth at $\tancoord$ & \\
 \eta &\!Integral over spatial phases, $\int\!\!d\tancoord\, \Phi(\tancoord)\mathcal{G}(\tancoord)$\!& \Eq~\ref{eq:single-scratch-otf-fourier-transform},~\ref{eq:spatial-phases-result}\\
 k,(k) &\!{}Scratch index (for ensemble summation)\!& \Eq~\ref{eq:total_amplitude}\\
 \hline
 \lambda & Optical wavelength & \\
 \mathcal{G} & Gaussian spatial filter & Fig.~\ref{fig:pixelzoom}, Sec.~\ref{sec:theory-coherence}\\
 \hline
 \mathcal{T}({\cvec{x}}) & Optical transfer function & Fig.~\ref{fig:angular-spectrum}b, Sec.~\ref{theory-spectral-brdf}\\
 {A_s} & {Shading area} & {\Eq~\ref{eq:diffracted-radiance-harvey}}\\
 \hline
 \end{tabular}
}
\caption{Overview of the notation used in this paper. The references point to the location where the quantity is introduced.}\label{table:notation}
\end{table}

\section{Preliminaries}
\label{sec:preliminaries}
Here we introduce the notation, geometric framework and theoretical background that serve as the foundation for our scattering model. An overview of notation can be found in Table~\ref{table:notation}.

\subsection{Scratch representation}\label{sec:preliminaries:scratch}
{
We represent a scratch as a curve $\scratchpos(\tancoord)$ parameterized by the arc length $\tancoord$, with local tangent vector $\uvec{t}(\tancoord) = {d{\scratchpos}(\tancoord)}/{d\tancoord}$. At any location $\tancoord$ along the scratch, surface normal $\uvec{n}(\tancoord)$, tangent $\uvec{t}(\tancoord)$ and bitangent $\uvec{b}(\tancoord)=\uvec{n}(\tancoord)\times\uvec{t}(\tancoord)$ form an orthonormal coordinate frame.
The geometric cross-section of the scratch at position $\tancoord$ is defined by the \emph{profile} $\mathcal{P}(\tancoord,b)$, which specifies a 1-dimensional height profile along the bitangential  coordinate b. Our model relies on scratch profiles that have analytical 1D Fourier transforms, such as rectangle or triangle shapes. Their scale in the $(\uvec{b}$-$\uvec{n})$-plane is given by the width and depth parameters $W(\tancoord)$ and $D(\tancoord)$, respectively (\Fig~\ref{fig:shading-geometry}).
} 
\subsection{Diffracted radiance}\label{sec:preliminaries:diffrad} Our model builds on tools from Fourier optics
\cite{Goodman1996}, specifically the \emph{angular spectrum} and the concept of
\emph{diffracted radiance} \cite{Harvey2000}, which we review here for
completeness. Being part of a \emph{scalar} theory of light transport, these
two tools assume that the electromagnetic field can be described by the
(scalar) amplitude of the oscillations that make up the electromagnetic field, as
opposed to the commonly used vectorial electric and magnetic fields. This
approximation is accurate in the far field and for diffracting apertures that
are larger than the wavelength of the radiation (both of these conditions are
satisfied by our application).
\begin{figure}[b]%
\centering
\includegraphics[width=.49\linewidth]{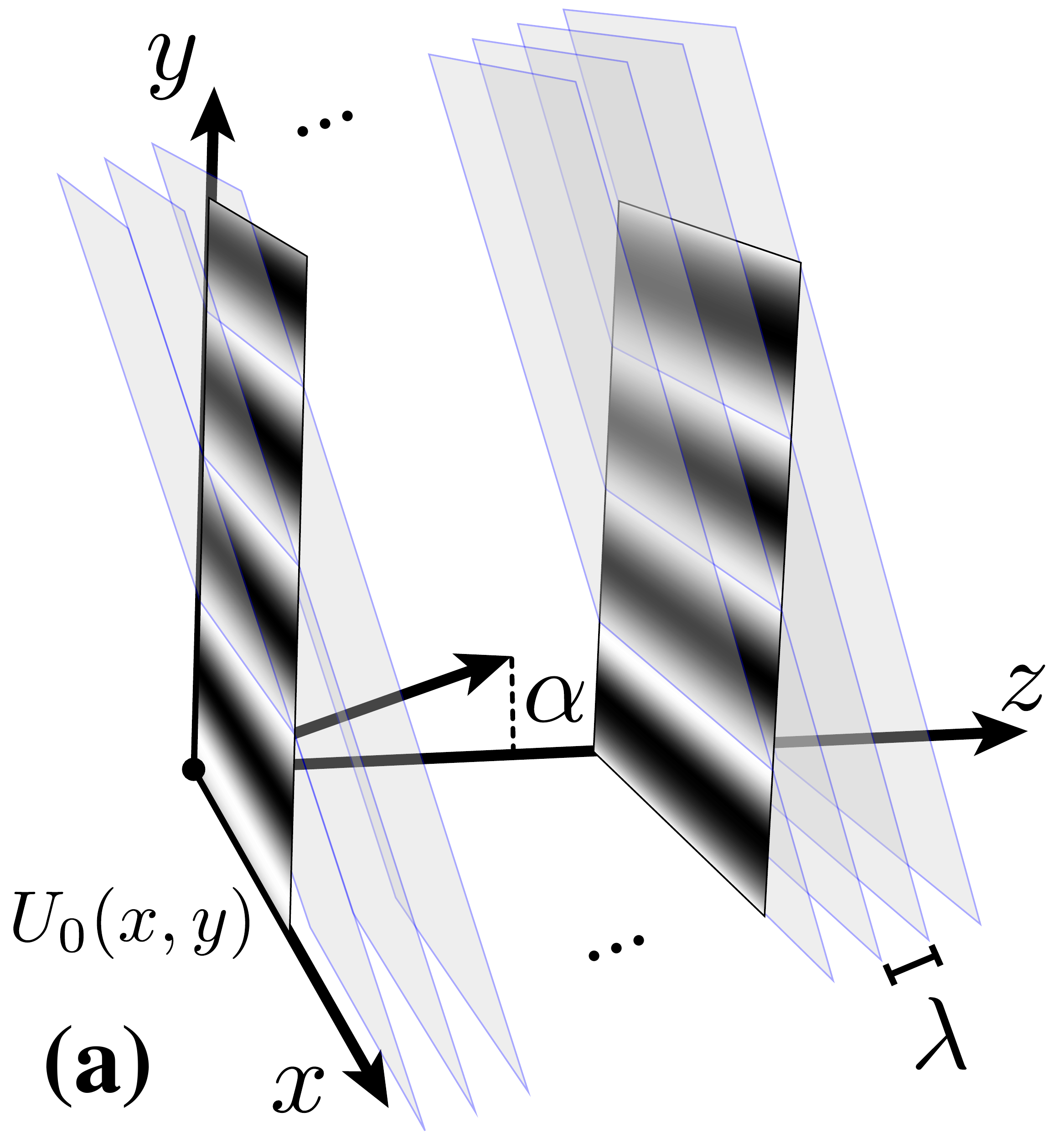}
\includegraphics[width=.49\linewidth]{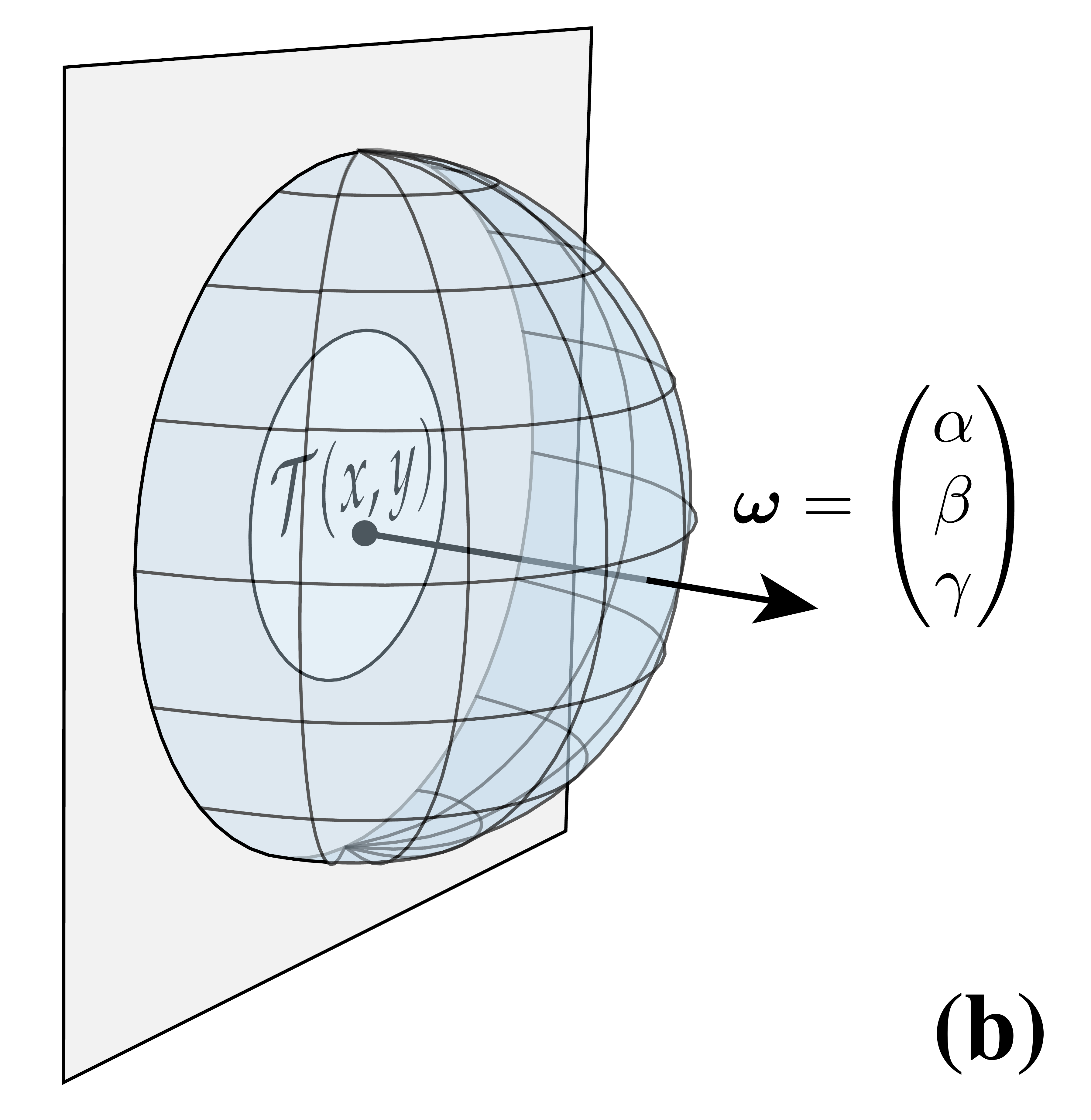}
    \caption{{(a),}
    Angular spectrum. Viewed along slices perpendicular to the $z$-axis, a
    monochromatic plane wave traveling in direction $(\alpha, 0,
    \sqrt{1\!-\!\alpha^2})$ causes vertical oscillations with frequency $\alpha/\lambda$;
    translating the slicing plane incurs a corresponding phase shift. Using the
    Fourier transform, this relation can be used to express an arbitrary field
    $U_0(x,y)$ incident at $z=0$ as a superposition of plane waves arriving
    from different directions. {(b)}, We are interested in the far-field
    diffracted radiance $L(\bm{\omega})$, which is proportional to the squared
    amplitude of the plane wave traveling in the same direction.
}
\label{fig:angular-spectrum}%
\end{figure}
Without loss of generality, we restrict ourselves to monochromatic radiation at
a wavelength of $\lambda$. The following discussion assumes that all
spatial coordinates are expressed in units of $\lambda$, since this leads to
simpler mathematical expressions.
Let $U(x, y, z)$ denote the scalar amplitude at position $(x, y,
z)^T$, and let $U_0(x, y)\coloneqq U(x, y, 0)$ denote a planar slice at
position $z=0$ (here called the aperture plane). A well-studied problem in this domain entails computing
$U(x,y,z)$ for $z>0$ given the amplitude distribution in the aperture plane $U_0(x, y)$. 
In the context of Fourier optics,
solutions can be found by taking the Fourier transform of all quantities in the
$xy$-plane, i.e.
\begin{equation}
    V(\alpha,\beta, z)\coloneqq\mathcal{F}\bigl\{U(\cdot, \cdot, z)\bigr\}_{\alpha, \beta},~
    V_0(\alpha,\beta)\coloneqq\mathcal{F}\bigl\{U_0(\cdot, \cdot)\bigr\}_{\alpha, \beta},
\end{equation}
and solving the Helmholtz equation $[\nabla^2+4\pi^2]U = 0$ analytically in
terms of the frequency-space representation $V$. The latter has an intuitive physical
interpretation: the amplitude $U(x, y, z)$ on any fixed $z$-slice can be described as a
superposition of plane waves arriving from different directions. In this context,
$V(\alpha, \beta, z)\in\mathbb{C}$ denotes both phase and amplitude of such a
plane wave arriving from direction $(\alpha, \beta, \gamma)$ where
$\gamma=\sqrt{1-\alpha^2-\beta^2}$ (Figure~\ref{fig:angular-spectrum}a).
The variables of this parameterization are referred to as \emph{direction cosines}.
Evaluating the superposition of plane waves is equivalent to an
inverse Fourier transform that recovers the original signal.
Assuming that radiation travels undisturbed through the half-space $z>0$, the
Helmholtz equation has a particularly simple solution which states that the
plane waves arriving at any $z$-slice correspond exactly to those at $z=0$
except for a phase shift $V(\alpha,\beta,z) = V_0(\alpha, \beta)e^{i 2\pi\gamma z}$. This solution
is exact under the stated assumptions, but the resulting field $U(x,
y, z)$ is prohibitively expensive to evaluate due to its definition in terms of
a pair of Fourier transforms. We instead rely on a far-field approximation,
which makes the reasonable assumption that the distance between the surface
and the camera is much greater than the wavelength of light 
(Figures~\ref{fig:pixelzoom}(d) and \ref{fig:angular-spectrum}(b)).

{This far-field approximation, known as \emph{diffracted radiance}, was
introduced by Harvey et al.~\shortcite{Harvey2000} and is defined as {
\begin{equation}
\label{eq:diffracted-radiance-harvey}
    L(\bm{\omega})= \frac{\lambda^2}{A_s} \bigl|\mathcal{F}\left\{U_0(\cdot, \cdot)\right\}\bigr|^2_{\alpha, \beta} = \frac{\lambda^2}{A_s}\bigl|V_0(\alpha, \beta)\bigr|^2, 
\end{equation}
where $\bm{\omega}=(\alpha,\beta,\gamma)$. $U_0$ describes both the source for the field at $z\!>\!0$ and the result of the radiation incident at $z\!=\!0$, as the corresponding angular spectrum $V_0(\alpha, \beta)$ is given by the superposition of plane waves from all directions.
A change of the direction of incident radiance by direction cosine $\beta_i$ results in a shift applied to all plane waves contributing to $V_0(\alpha, \beta)$, and the angle-shifted angular spectrum now reads $V_0(\alpha, \beta\!-\!\beta_i)$ (we show the one-dimensional case for simplicity, but the concept holds for the second dimension as well). As angular spectrum and complex amplitude are related by a Fourier transform, this can be 
interpreted in terms of the Fourier shift theorem as a linear phase shift applied to $U_0$. 
An additional attenuation factor, the third direction cosine $\gamma_i$~\cite{Harvey2000}, accounts for the decreased intensity at oblique incident angles and modulates \Eq~\ref{eq:diffracted-radiance-harvey} as
\begin{eqnarray}
\label{eq:diffracted-radiance-derivation}
    L(\bm{\omega}, \alpha_i, \beta_i)\!&=&\!\gamma_i \frac{\lambda^2}{A_s}\bigl|V_0(\alpha\!-\!\alpha_i, \beta\!-\!\beta_i)\bigr|^2\!\\
    &=&\!\gamma_i \frac{\lambda^2}{A_s} \bigl|\mathcal{F}\bigl\{U_0(\cdot, \cdot)\,e^{2\pi i \left(\beta_i {y} + \alpha_i {x}\right)}\bigr\}_{\alpha, \beta} \bigr|^2\!\!.
\end{eqnarray}
}
This influence of the angular distance in direction cosine space on the diffracted radiance is also known as \emph{shift invariance}.}
\subsection{BRDF model\label{theory-spectral-brdf}} To quantify the interaction of light with a surface exhibiting microscale defects in physically-based rendering frameworks, we first start with the well-known definition of the bidirectional reflectance distribution function (BRDF) 
\begin{equation}
 f_r = \frac{dL_\mathrm{s}({\cvec{x}},{\uvec{\omega}_o})}{dE_i({\uvec{\omega}_i})},
 \label{eq:general-bsdf}
\end{equation}
which relates differential irradiance to scattered radiance. Here, $\cvec{x}$ represents a position on the surface, $\uvec{\omega}_i$ the direction from which this surface is illuminated and $\uvec{\omega}_o$ the direction from which it is observed. 

The radiance scattered by a diffracting aperture is given by \Eq~\ref{eq:diffracted-radiance-derivation} as a function in direction-cosine space 
using a coordinate system where all spatial variables are normalized to the optical wavelength. 
A change of variables enables us to rewrite the representation of the Fourier transform in a non-scaled coordinate system as
\begin{equation}
 \label{eq:diffracted-radiance}
    L_s(\cvec{\xi}) = \gamma_i \frac{1}{A_s} \frac{1}{\lambda^2} \bigl|\mathcal{F}\bigl\{ U_0(\cvec{x})\bigr\} \bigr|^2_{\cvec{\xi}}
  \end{equation}
  {
with 
\begin{equation}
 \cvec{\xi} = \frac{1}{\lambda} \cdot
	    \begin{pmatrix}
                     \alpha_o - \alpha_i\\
                     \beta_o - \beta_i\\
                     \gamma_o - \gamma_i
            \end{pmatrix}
\end{equation}
}%
Utilizing a first order Born approximation, similar to Church and Takacs~\shortcite{Church1995}, 
we can describe the complex wavefront $U_0(\cvec{x})$ in the surface plane by the modulation of the wavefront of the incident light $U_\incident(\cvec{x})$ with the so-called 
\emph{transfer function} $\mathcal{T}(\cvec{x})$ \cite{Lipson2010,Goodman1996} of the diffracting plane as
\begin{equation}
\label{eq:born-approx}
  U_0(\cvec{x}) = U_\incident(\cvec{x}) \cdot \mathcal{T}(\cvec{x}).
\end{equation}
Since the diffracting aperture is uniformly illuminated (\Sec~\ref{sec:preliminaries}), we can neglect the position dependence of the complex amplitude of the incident 
light in the aperture plane. Thus, $U_\incident(\cvec{x}) = U_\incident$ is a constant modulation factor. 
Substitution into \Eq~\ref{eq:diffracted-radiance} then yields
 \begin{equation}
   L_s(\cvec{\xi}) = \gamma_i \frac{1}{A_s} \frac{1}{\lambda^2} \left|U_\incident\right|^2 \bigl|\mathcal{F}\bigl\{ \mathcal{T}(\cvec{x})\bigr\}_{\cvec{\xi}} \bigr|^2 
\end{equation}
where $E_\incident = |U_\incident|^2$ is the incident irradiance. 
{Applying \Eq~\ref{eq:general-bsdf}, we can therefore write}
\begin{equation}
 \label{eq:bsdf_general}
 f_r(\cvec{\xi}) = \gamma_i\frac{1}{A_s} \frac{1}{\lambda^2} \bigl|\mathcal{F}\bigl\{ \mathcal{T}(\cvec{x})\bigr\}_{\cvec{\xi}} \bigr|^2 
\end{equation}
which is the non-paraxial spectral BRDF for reflected light that is diffracted by a surface exhibiting microscale features represented by an 
optical transfer function $\mathcal{T}(\cvec{x})$. 
{
Diffracted radiance shares obvious similarities with Stam's general BRDF \cite[\Eq~7]{Stam1999}. However, both rely on different formulations as Stam's model explicitly utilizes Kirchof theory (tangent-plane approximation, Huygens principle) whereas 
diffracted radiance can be solely derived from the angular spectrum of plane waves which is a linear systems approach to the same problem~\cite{Krywonos2006} and, in a non-simplified form, leads to the more rigorous Rayleigh-Sommerfeld diffraction formalism.
}
%
\begin{table}[t]
\centering
{\small\begin{tabular}[T]{|l|r|}
\hline
 {\textbf{Assumption}} & {\bf{Reference}}\\
 \hline
 {Diffracted radiance} & \\
 {Far-field scattering} & \Sec~\ref{sec:preliminaries:diffrad}; \Eq~\ref{eq:diffracted-radiance-harvey},~\ref{eq:bsdf_general}\\
 \qquad{First order Born approx.} & \Eq~\ref{eq:born-approx}\\
 \hline
 \hline
 {Spatial coherence} & \\
 \qquad{Equal coherence condition for the full scene} & \Sec~\ref{sec:theory-coherence}\\
 \qquad{Intensity drops of towards edges of a light source} & \Sec~\ref{sec:theory-coherence}\\
 \hline
 \hline
 {Surface representation} & \\
 \qquad{Homogeneous base material} & \Sec~\ref{sec:theory-scratch-brdf:composition}; \Eq~\ref{eq:total_amplitude}\\
 \hline
 \hline
 {Scratches} & \\
 \qquad{No intersections} & \Eq~\ref{eq:total_amplitude}\\
 \qquad{No self-shadowing/masking} & --\\
 \qquad{Linear segments with constant profile} & \Eq~\ref{eq:single-scratch-otf-fourier-transform}\\
 \qquad{Spatial--spectral separability} & \Sec~\ref{sec:theory:singlescratch}; \Eq~\ref{eq:single-scratch-otf-fourier-transform}\\
 \hline
 \end{tabular}
}
\caption{{References to assumptions used by our model}}\label{table:assumptions}
\end{table}
%
{
\section{A diffraction SVBRDF for scratched surfaces}\label{sec:theory-scratch-brdf}
The concept of ray tracing is fundamentally incompatible with the basic principles of wave optics. 
In fact, the wave-optical counterpart of a ``ray'' with sharply defined direction is a plane wave of infinite lateral extent 
{and equal amplitude at each point on the sufficiently far away illuminated surface. Following \Eq~\ref{eq:bsdf_general} this leads to a Fourier transform of the whole surface, regardless of its extent, 
which would not allow us to resolve single surface features and is only correct for point-light sources. To provide a well-founded condition for mutual interference between scratches and re-introduce the ability 
to resolve scratches, we draw on coherence theory to develop a physically justified interface between the long-range (ray-optical) light transport within a path tracing framework and the wave-optical scattering model. 
To this end, we will first introduce the concept of spatial coherence which allows us to define a coherence window and leads to a spatially varying BRDF (SVBRDF. 
We will then study the wave-optical contribution of a single scratch in isolation, and finally look at large ensembles of scratches.}
\subsection{Spatial coherence - resolving scratches}\label{sec:theory-coherence}
\begin{figure}[t]
\centering
\includegraphics[width=0.7\linewidth]{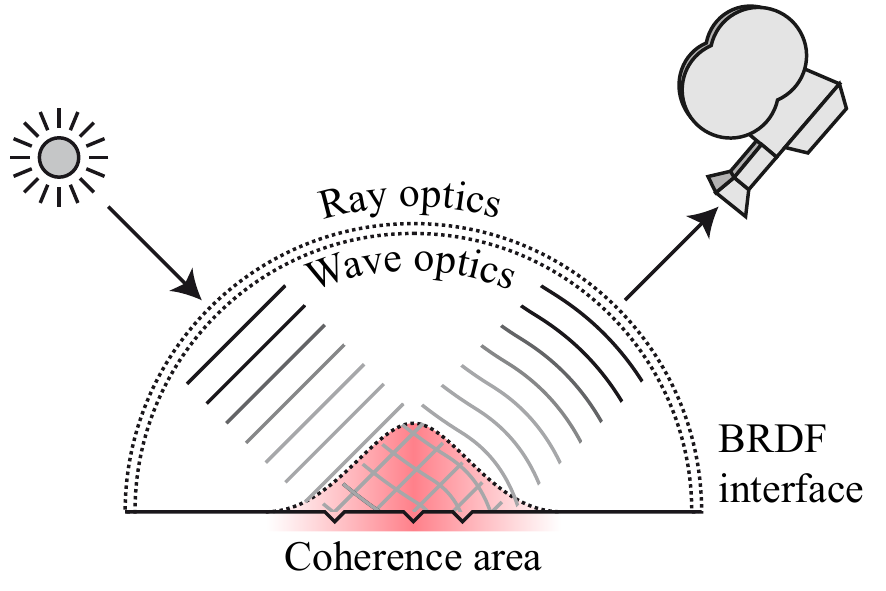}
\caption{To be compatible with a standard global illumination renderer, our model constrains wave optics to the reflectance model. 
The coherence area, represented by the Gaussian filter $\mathcal{G}$, marks the portion of the surface where scattered light waves will superimpose coherently. 
The interface to the outside world is provided in terms of geometric optics and radiance units. The computation of the coherence area is explained in \Sec~\ref{sec:theory-coherence}}%
\label{fig:rayoptics-interface}%
\end{figure}
The van Cittert–Zernike theorem relates the angular extent of a light source illuminating a surface with a spatial filter on the surface via a Fourier transform~\cite{Goodman1996,Lipson2010}. 
{For instance, a point-light source yields a constant infinite spectrum whereas a disk-shaped area light would result in an Airy function. This \emph{coherence function} 
defines the corresponding spatial weights. More intuitively, it defines which structures in the vicinity of the observed point, such as scratches, actually contribute to the wave-optical scattering that leads to 
diffraction and interference. The distance between the first zero-crossings of the coherence function usually is used to define the \emph{coherence area}, the extent of the filter. For example, a circular star 
of diameter $d = 0.07$ arcsec has a coherence area whose radius is $r_c = 1.22\lambda/d\approx\,1.9$m~\cite{Lipson2010}. After this first zero-crossing, the coherence function may predict further areas of coherence 
which are suppressed if the light source is less intense around its' edges.
The concept of spatial coherence is the key to our goal of resolving scratches in a wave-optical scattering model; being inherently interwoven with wave-optics it allows us to 'select' and resolve scratches
with a coherence-based resolution limit.}
{
To approximate the coherence function, we make two assumptions: First, we assume the coherence condition to be equal everywhere in the considered scene, which allows us to define a single coherence function. 
Second, we will neglect coherence effects outside of the central peak, inherently assuming a fall-off of the light-source intensity towards its' corners. More formally, we define the coherence function 
as a spatial kernel given by $\mathcal{G}(\cvec{x} - \cvec{x}_0)$, controlling the relative weight of points on the surface with respect to the intersection point $\cvec{x}_0$ of the ray on the surface, 
similar to \cite{Dhillon2014}. However, in contrast to our model which utilizes this concept to re-enable resolving surface structure, 
\cite{Dhillon2014},due to their finite measurement size, mainly suppress Fourier transform artifacts caused by discretization. Here, the entire object surface is covered in scratches, 
and the filter kernel forms a windowing function that determines which subset will contribute to the light received by the pixel, similar to a short-time Fourier transform.
We assume the spatial kernel \begin{equation}\mathcal{G}(\cvec{x}) = e^{-\frac{1}{2}{|\cvec{x} - \cvec{x}_0|^2}/{\sigma^2}} \end{equation}
to be an isotropic Gaussian in the tangent plane at any point of intersection (see \Fig~\ref{fig:rayoptics-interface}), thus representing the central peak of typical coherence functions for finite light sources. 
We define the coherence areas' diameter associated with a Gaussian to be given by $\delta_c = 6\sigma$ as almost all information is contained within this interval.
}
Thus, we modify \Eq~\ref{eq:bsdf_general} to include spatial coherence via 
\begin{equation}
 \label{eq:bsdf_coherence}
 f_r(\cvec{x}_0,\cvec{\xi}) = \gamma_i\frac{1}{A_s} \frac{1}{\lambda^2} \bigl|\mathcal{F}\bigl\{ \mathcal{T}(\cvec{x}) \cdot \mathcal{G}(\cvec{x} - \cvec{x}_0)\bigr\}_{\cvec{\xi}} \bigr|^2
\end{equation}
{
A common literature value for $\delta_c$ for scenes under direct illumination by sunlight or a light bulb in a kitchen environment \cite{MandelWolf1995,Divitt2015} is $\delta_c = \SI{60}{\micro\meter}$ which 
we choose to approximate such illumination situations. Note that  \Eq~\ref{eq:bsdf_coherence} closely reproduces the concept of spatial coherence: For a point-light source, $\delta_c \rightarrow \infty$ and 
$\mathcal{G}(\cvec{x}) \rightarrow const.$ which reduces the SVBRDF to the BRDF of \Eq~\ref{eq:bsdf_general} and thus does not allow to resolve surface features. On the other hand, an infinitely extended light source 
would yield $\delta_c \rightarrow 0$ and we obtain a dirac-delta for the coherence function which is in full accordance to optics literature. Additionally, the full reflectance a pixel receives is given not by 
\Eq~\ref{eq:bsdf_coherence} but by its' value integrated over the pixel footprint so that $\cvec{x}_0 \in \mathrm{Footprint}$\cite{Levin2013}. Intuitively, the coherent subsamples are thus first converted to 
radiance and then averaged. Fortunately, a ray tracer inherently samples the pixel footprint and we state in \Sec~\ref{sec:implementation} how we can exploit the coherence area for performance improvements to counter this numerical integration. Still, our 
model a priori requires more samples per pixel than standard BRDFs to account for spatial variation. 
}
%
\subsection{Additive composition of transfer function}\label{sec:theory-scratch-brdf:composition}
The key part of our BRDF is the Fourier transform of the optical transfer function $\mathcal{T}(\cvec{x})$ which, in general,
can be written as
\begin{equation}
 \label{eq:general_otf}
 \mathcal{T}(\cvec{x}) = A(\cvec{x}) \cdot e^{i \phi(\cvec{x})} = {A'(\cvec{x}) \sqrt{F(\cvec{x})} \cdot e^{i {4 \pi h(\cvec{x})}/{\lambda} }}
\end{equation}
where $A = A' \sqrt{F}$ is the amplitude factor of the surface material, $F$ is the Fresnel factor and $\phi(\cvec{x}) = 4 \pi h(\cvec{x})/\lambda$ is the change of phase induced by the height variation of the microstructure. 
{We now apply this concept to scratched surfaces where each scratch is described by its individual transfer function resembling which introduces the aforementioned phase changes by locally altering the height 
of the scratched object.}
}

{
We assume our materials to consist of a homogeneous \emph{base material} that does not exhibit spatial variation. This {base material} exhibits {defects} (our scratches) 
at distinct positions, allowing us to redefine our transfer function as a {base transfer function} 
from which we subtract \emph{masks} covering the defects and adding the {defects} back at the same position. 
This implies that masks and scratches must cover the same area on the surface. More formally, 
\begin{equation}
\label{eq:total_amplitude}
\mathcal{T}(\cvec{x}) = \mathcal{T}_\mathrm{base}(\cvec{x}) - \sum_k \mathcal{T}^{(k)}_{\mask}(\cvec{x}) + \sum_k \mathcal{T}^{(k)}_{\scratch}(\cvec{x}),
\end{equation}
where the superscript $(k)$ denotes the $k^\textrm{th}$ mask-scratch pair and $\mathcal{T}_\mathrm{base}(\cvec{x}) 
= \mathcal{T}_\mathrm{base}=A_\mathrm{base} \sqrt{F_\mathrm{base}}$. 
This decomposition of surface structure into individual scratches, and the additive superposition of their contributions, is the key to a practical implementation of our model, since it allows for efficient analytical evaluation of the Fourier transform in \Eq~\ref{eq:bsdf_general}. 
However, it implies two simplifications that have to be considered. Firstly, we assumed the base to be a perfectly flat mirror; hence, any form of surface roughness would have to be emulated by an intractably dense distribution of scratches. As a workaround, in \Sec~\ref{sec:implementation} we introduce a way of blending our model with multifacet BRDFs, where the area coverage of scratches acts as blending weight. Secondly, the model overestimates intersection regions. In a region where two scratches overlap, the base contribution will be subtracted (masked) twice and replaced by the sum of two scratches. As we show in \Sec~\ref{sec:results}, the effect of this approximative handling of intersections on the rendered outcome is quite minor. Note that as long as scratches and base have the same amplitude factor, this procedure only affects the  diffracted phase but does not violate energy conservation. 
To provide a quick reference, all the assumptions and simplifications used for our model can be found in Table~\ref{table:assumptions}.
\subsection{Single-scratch transfer function}\label{sec:theory:singlescratch}
The linearity of the Fourier transform allows us to first consider a single scratch transfer function (or its transform) and later extend the concept to the full solution. The local geometry of each scratch is defined by the profile $\mathcal{P}(\tancoord,b)$ (\Sec~\ref{sec:preliminaries:scratch}, \Fig~\ref{fig:shading-geometry}), which is a 1-dimensional height distribution across the scratch defined as a function of the bitangent coordinate $b$. 
This results in a scratch-space optical transfer function which is defined in local coordinates: 
\begin{align}
\label{eq:profile_transfer_function}
\mathcal{T}_{\mathrm{scratch}}(b,\tancoord)  &= 
A'(b,\tancoord) \sqrt{F(b,\tancoord)} \cdot e^{i {4 \pi \mathcal{P}(b,\tancoord)}/{\lambda}} .
\end{align}
The integration of all transfer functions along a scratch yields a full spatial representation of the amplitude and phase changes induced by the material itself and the height variations. 
\paragraph{Incorporating the scratch profile. }
{
Intuitively, the Fourier transform of a scratch transfer function can be understood as the integral over the Fourier transform of the rotated and shifted 1d-transfer functions along the scratch trajectory whose 
complexity is determined by profile variation and curvature of the scratch.
}
For simplicity, we consider the scratches to consist of linear scratch segments whose profile does not change along the scratch. This can be done without the loss of generality as arbitrary curves of varying profile 
can always be split into linear segments of constant profile (and thus constant transfer function). 
{
In addition, we assume only the spatial phases to be affected by the filter, i.e., the width of a scratch is negligible in comparison to its length with respect to the coherence area
This allows us to separate the spatial (position) and spectral (profile) components of each scratch and to express its' transfer function in the scratches' 
own tangent space with axes $\uvec{t}$, $\uvec{b}$ and $\uvec{z}$. The Fourier transform of a single scratch transfer function then reads
\begin{eqnarray}
\label{eq:single-scratch-otf-fourier-transform}
\mkern-40mu&&\mathcal{F}\left\{\mathcal{T}_{\mathrm{scratch}}(b,\tancoord)\right\}_{\cvec{\xi}'} \approx \nonumber\\
\mkern-40mu&&\mathcal{F}\left\{ \mathcal{T}_{\mathrm{scratch}}(b)\right\}_{\cvec{\xi}_2} \cdot \!\int\!\!d\tancoord\,\mathcal{G}(\tancoord)\,\Phi(\tancoord) = \\\nonumber
\mkern-40mu&&\!\int\!\!db\,\mathcal{T}_{\mathrm{scratch}}(b) e^{-2\pi i b\uvec{r}_{r,2}'\cvec{\xi}_2'}\,\!\int\!\!d\tancoord \,e^{-|\cvec{r}_{r}'(\tancoord)|^2/(2\sigma^2)}\,e^{-2 \pi i (\cvec{r}_{r}'(\tancoord)\cdot\cvec{\xi}')}
\end{eqnarray}
where the projection into tangent space is given by the inverse of the rotation matrix defining the orientation of the scratch so that
{
\newcommand*{\horzbar}{\rule[.5ex]{1.8ex}{0.5pt}}
\begin{equation}
\label{eq:scratch-frame-trafo}
 \cvec{x}' = \cvec{R}^{-1} \cvec{x} = 
 \left[
  \begin{array}{ccc}
    \horzbar & \uvec{t}^{T}(\tancoord) & \horzbar \\
    \horzbar & \uvec{b}^{T}(\tancoord) & \horzbar \\
    \horzbar & \uvec{z}^{T} & \horzbar
  \end{array}
\right]	    \cvec{x}\,;\, \cvec{r}' = \cvec{R}^{-1}(\cvec{r} - \cvec{x}_0) 
 \end{equation}}
}
and $\cvec{r}'$ is the relative scratch position. With \Eq~\ref{eq:single-scratch-otf-fourier-transform} at hand, we are now able to express scratches of in principle arbitrary profile in terms of an optical transfer 
function which can be used to express the corresponding diffraction effects.
}
\subsection{Scratch ensemble solution}
With \Eq~\ref{eq:single-scratch-otf-fourier-transform} we are now able to express the Fourier transform of a single scratch, 
{
allows us to extend the solution to scratch ensembles. 
}
In addition, scratches and masks share the same spatial information per construction, thus the 
integral of the weighted spatial phases of scratch $(k)$ is equal to the corresponding integral of mask $(k)$, mask and scratch only differ in the respective profile.
In other words, the Fourier transforms of a mask-scratch pair have the same spatial phases $\mathcal{G}(t)\Phi(t)$. Recalling that we use a homogeneous material (Fresnel factor not spatially varying) and 
substituting \Eq~\ref{eq:single-scratch-otf-fourier-transform} into \Eq~\ref{eq:bsdf_coherence}
{
we can intuitively split the Fourier transform into base- and scratch related terms to obtain
}
\begin{equation}
  f_r(\cvec{x}_0,\cvec{\xi}) = \gamma_i \frac{F}{\pi \sigma^2} \frac{1}{\lambda^2}
  \left|\mathcal{B}(\cvec{\xi}) -  \mathcal{S}(\cvec{\xi}_2')\right|^2
  \label{eq:scratch_general_brdf}
 \end{equation}
{
where $A_s = \pi \sigma^2$ is the area under the squared amplitude of the Gaussian (shading area) and $F$ the Fresnel coefficient of the homogeneous material. We define 
\begin{equation}
 \mathcal{B}(\cvec{\xi}) = A'_\mathrm{base} 2\pi \sigma^2 e^{-2 \pi^2 (\xi_1^2 + \xi_2^2)/\sigma^2},
 \label{eq:base-response}
\end{equation}
as the \emph{base response} given by the Fourier transform of the filter kernel and resembling the undisturbed reflectance of the material without scratches. On the other hand, the \emph{scratch response}
\begin{equation}
\mathcal{S}(\cvec{\xi}_2') = \sum_k \left[\mathcal{F}\left\{\mathcal{T}^{(k)}_{\mathrm{mask}}\right\}_{\cvec{\xi}_2'} - \mathcal{F}\left\{\mathcal{T}^{(k)}_{\scratch}  \right\}_{\cvec{\xi}_2'}\right] \eta^{(k)}(\cvec{x}_0, \cvec{\xi}').
 \label{eq:scratch-response}
\end{equation}
then defines the disturbance of the smooth heightfield by scratches. It only relies on the scratches' profiles and their location on the surface with resepect to the point of intersection encoded into the 
integral over the spatial phases $\eta^{(k)}$ (see \App~\ref{app:spatial-phases} for the full solution).
In its simplest form using a rectangular scratch profile (see \App~\ref{app:profiles}) and $A'_\mask = A'_\scratch = A'_\base$, the scratch response function is
\begin{eqnarray}
\mkern-40mu&&\mathcal{S}(\cvec{\xi}_2') = A'_\mathrm{base} \sum_k \mathcal{W}^{(k)} \mathcal{D}^{(k)}\eta^{(k)}(\cvec{x}_0, \cvec{\xi}')\\
\mkern-40mu&&\mathcal{W}^{(k)} = W^{(k)} \mathrm{sinc}\left(\pi \frac{W^{(k)}}{\lambda} \cvec{\xi}'_2\right)\\
\mkern-40mu&&\mathcal{D}^{(k)} = \left( 1 - e^{4\pi i {D^{(k)}}/{\lambda}}\right)\\
 \label{eq:ex-scratch-response}
\end{eqnarray}
where we further separate the dependence of the individual diffraction patterns of the scratches on the respective width and depth via the width-term $\mathcal{W}^{(k)}$ and the corresponding depth term $\mathcal{D}^{(k)}$. 
This forms our reflectance function for rendering surfaces with microscale scratches. In \Sec~\ref{sec:implementation}, we explain how our rendering system efficiently looks up the scratches that are relevant for a given shading event.
}
%
%
\begin{figure*}[t]%
\centering
\includegraphics[width=0.33\linewidth]{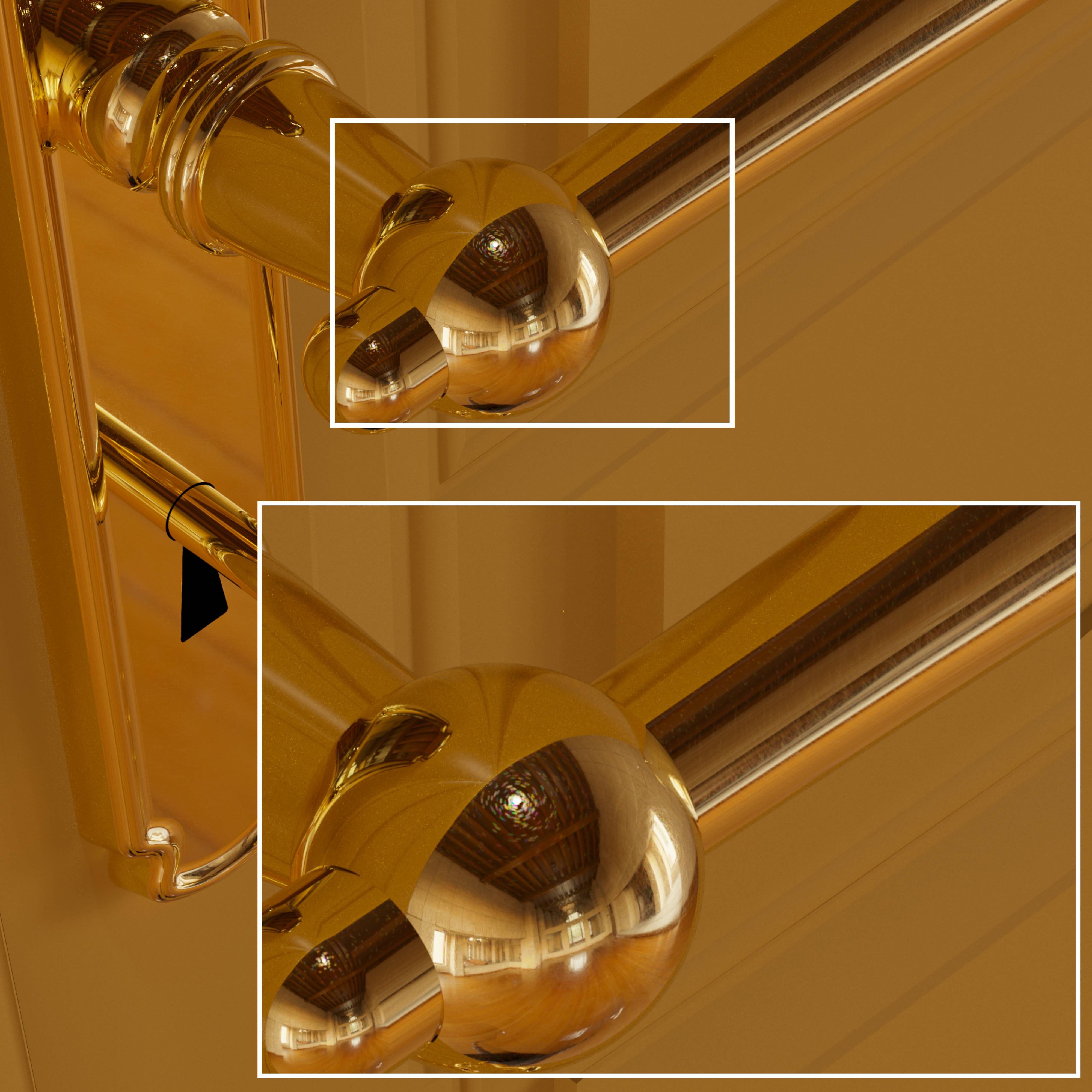}\hfill%
\includegraphics[width=0.33\linewidth]{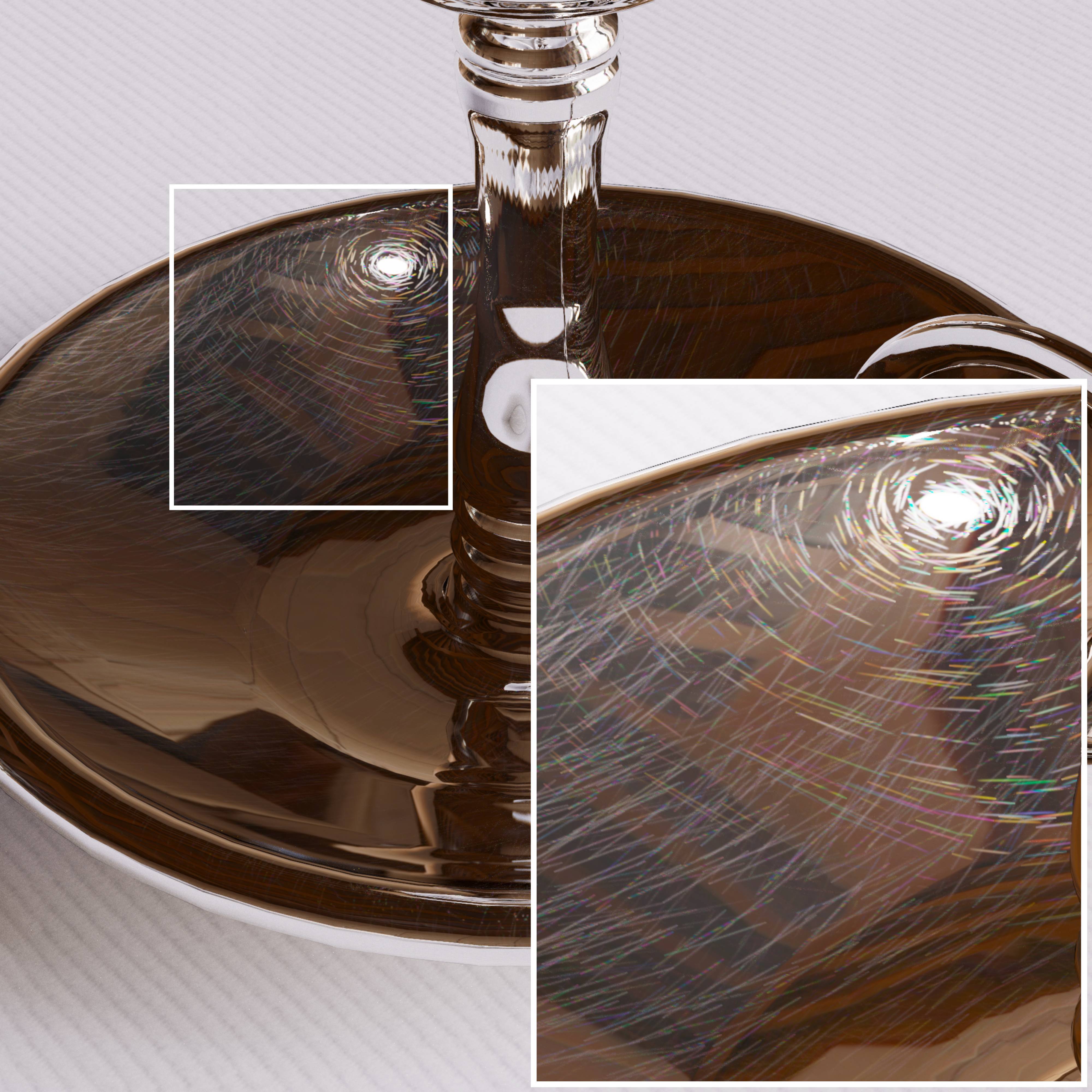}\hfill%
\includegraphics[width=0.33\linewidth]{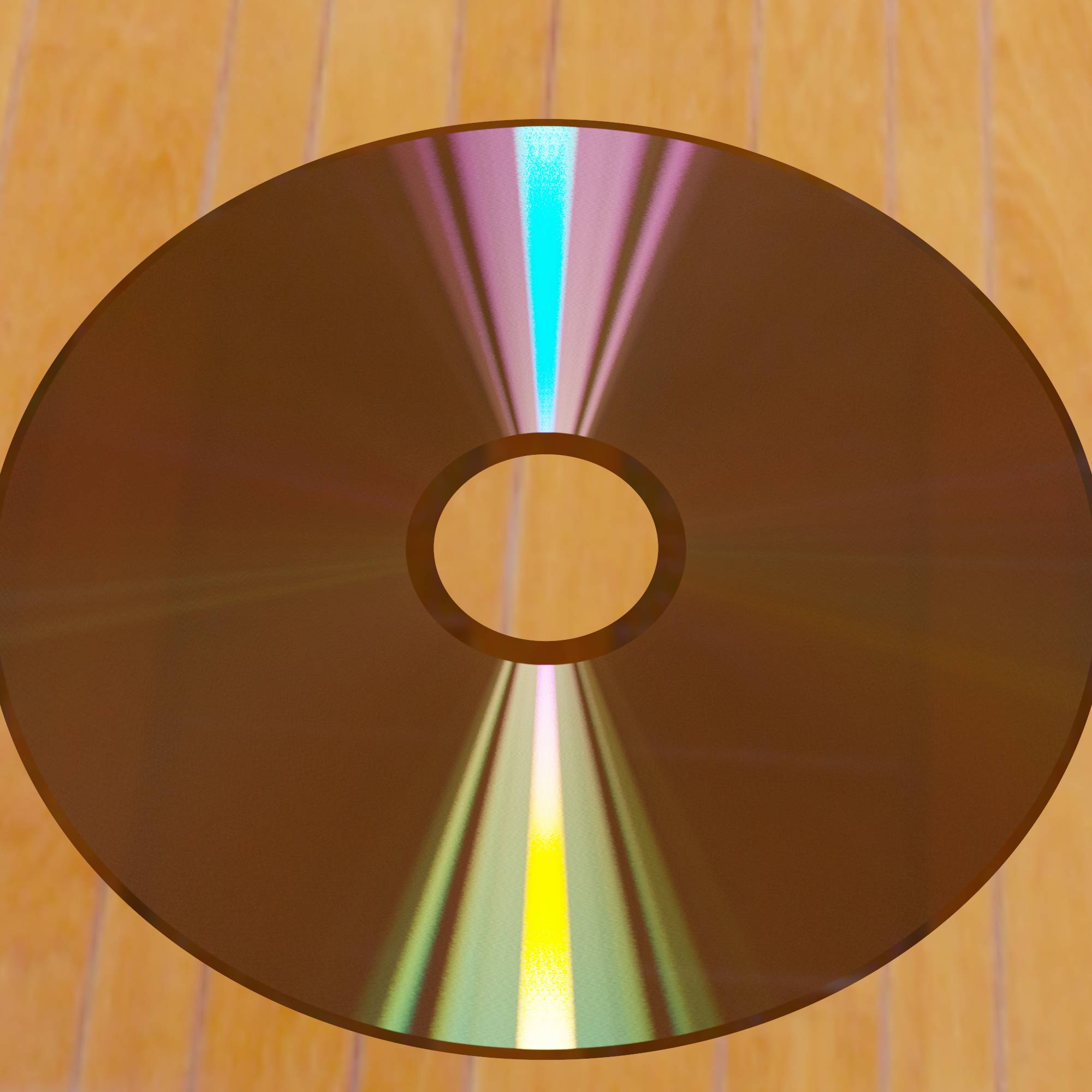}\\
\includegraphics[width=\linewidth]{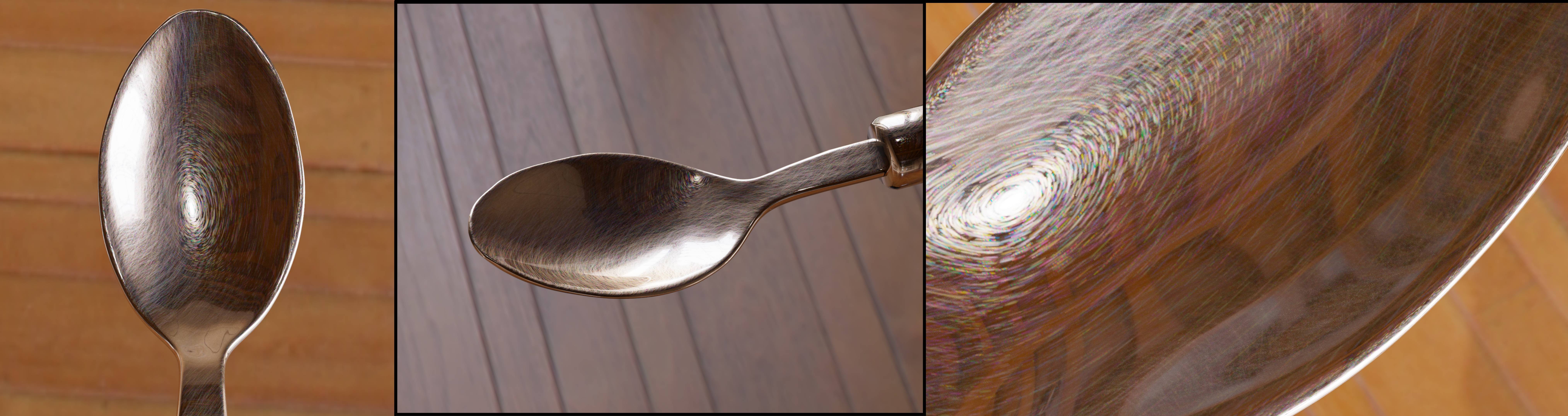}\hfill
\caption{We demonstrate our reflectance model on four example scenes. Top row from left to right: A golden door handle, a candle holder, both with randomized 
scratch distributions. A compact disc with circular scratches with constant separation. Bottom row: A spoon from different view points, the scratches 
are randomized. The top-down view reveals the scratches with iridescent colors that lie on a circle around the specular highlight. In the side-view 
the specular highlight only subtends a small fraction on the spoon and under incoherent illumination by the surrounding lights the scratches appear mostly white. 
At close-up again color as well as geometry of the scratches is revealed. \nonarXiv{We present complementary videos for these scenes in the supplemental material.}}
\label{fig:scenes}%
\end{figure*}
\section{Usage in a rendering framework}\label{sec:implementation}
In \Sec~\ref{sec:theory-scratch-brdf}, we derived a BRDF for surfaces with micro-scale scratches that is compatible with standard ray tracing-based rendering systems. We use a path tracer that evaluates the BRDF at intersections of rays, cast from a camera, with a surface to which the BRDF is bound. The scratches are either 
applied to the surface by defining positions and directions as well as scratch parameters directly, usually by drawing from a distribution, or by scratching an arbitrary mesh 
using an editing tool. {In full-spectral rendering mode (all figures except \Fig~\ref{fig:pattern-mapping}), the renderer samples 16 wavelengths across the visible range. 
A reduced RGB version represents the primary colors by the wavelengths $\lambda_\mathrm{red} = 700\,\mathrm{nm}$; $\lambda_\mathrm{green} = 520\,\mathrm{nm}$ and $\lambda_\mathrm{blue} = 440\,\mathrm{nm}$.
}
\paragraph{Scratch data structure and lookup}
{
The scratch particles are represented by line segments, which we store in a bounding volume hierarchy (BVH) consisting of axis-aligned bounding boxes (AABB). The BVH is build using sorting on a space-filling curve (Morton code builder~\cite{Lauterbach2009}) and efficient traversal is ensured by employing 
the skip-pointer structure proposed by~\cite{Smits1998}.
To reduce spatial overlap between the elements of this structure, we perform further splits. Since the shading cost of our model generally outweighs the 
intersection test cost, we do not directly split the scratch particles. We instead use a directed acyclic graph structure where multiple AABBs can 
be associated with the same leaf element. During intersection testing we store only the unique intersections within the shaded area. 
When shading a point on the surface, we only consider scratches within the pixel footprint. As stated in \Sec\ref{sec:theory-coherence}, we need to perform an integration over the pixel footprint to achieve 
the correct incoherent superposition of the coherent subsamples. Rays that strike the surface at an oblique angle might query many scratch particles, leading to poor performance due to a large 
pixel footprint. At this point however, we know the spatial weight each scratch that would be queried would achieve. We can thus improve performance by not intersecting the BVH with the pixel footprint but 
with a sphere whose diameter matches the extent of the Gaussian filter kernel, because scratches outside only have a negligible contribution. To this end we set this diameter to $\Delta_c = 6\sigma= 60\mu m$ and 
only query scratches that actually contribute.
}
\paragraph{Importance sampling the base response function}
To importance sample the base surface response \eqref{eq:base-response}, we
generate two normally distributed samples and 
{
scale them by the standard deviations of the (Gaussian) target distribution in the angular spectrum, 
resulting in a sampled angular frequency $\bm{\xi}$. Specifically, we set
{
\begin{equation}
\cvec{\xi} = [(\sqrt{8}\pi \sigma_p)^{-1}, (\sqrt{8}\pi \sigma_s)^{-1}]^{T}.
\end{equation}
}
Next, the sampled frequency is used to map the incident direction onto a
scattered direction $\uvec{\omega}_o=(\alpha_o, \beta_o, \gamma_o)^T$ by
solving for $\alpha_o$ and $\beta_o$ via the shift invariance property of
diffracted radiance, i.e.~$\alpha_i+\alpha_o=\xi_1$ and
$\beta_i+\beta_o=\xi_2$, to obtain
{
\begin{equation}
  \uvec{\omega}_o = 
		  \begin{pmatrix}
		    \alpha_o\\
		    \beta_o\\
		    \gamma_o
                  \end{pmatrix}
                  = 
                  \begin{pmatrix}
		    \xi_1 - \alpha_i\\
		    \xi_2 - \beta_i\\
		    \sqrt{1-\alpha_o^2-\beta_o^2}
                  \end{pmatrix}  
\end{equation}
}
}
Two details must be noted regarding this step: occasionally, a sample satisfies
$1-\alpha_o^2-\beta_o^2<0$, which does not lead to a valid scattered direction.
These samples correspond to evanescent waves that do not propagate, and the
associated sample is simply dropped.
Secondly, sampling a position in the angular frequency domain and mapping it on
the outgoing hemisphere corresponds to a change of variables that appears both
in the sampling density and Monte Carlo weight of this sampling strategy. The
mapping is simply the parallel projection from the unit disc to the unit
hemisphere known as the Nusselt analog, and the Jacobian determinant {prefactor for the Gaussian PDF} associated
with this mapping is the {direction} cosine $\gamma_o$.
\paragraph{Importance sampling of the scratch response function}
Importance sampling of the scratch response relies on a modification of a
sampling technique that was originally developed by d'Eon et al.~\shortcite{deon2011energy} in the context of hair rendering. Given an incident
direction ($\phi_i, \theta_i$) expressed in the coordinate system of a hair
fiber, this technique works by sampling a specular reflection from an ideally
reflecting cylinder, producing a reflected direction on a Dirac delta circle of
azimuths with elevation angle $\theta_i=-\theta_o$. To account for roughness,
the direction $(\phi_o, \theta_o)$ is then perturbed by a random offset drawn
from a spherical von  Mises-Fisher distribution with concentration parameter $\kappa$.
The spherical density of sampled directions has an explicit form in terms of
a modified Bessel function of the first kind, specifically
\[
p(\theta_o,\phi_o) = 
\frac{
\kappa
}{4\pi \sinh\kappa}
e^{-\kappa\cos\theta_i\cos\theta_o}
I_0\left[
\kappa\sin\theta_i\sin\theta_o
\right]
\]
{
Although disconnected from the explicit profiles of scratches}, we found the resulting 
distribution to be an excellent match for the response
function of individual scratches when interpreting the scratch tangent vector
as a fiber direction and mirroring reflected directions that would enter the
surface along the normal direction, doubling the density $p(\theta_o,\phi_o)$
for directions that lie in the upper hemisphere.

Our method applies multiple importance sampling via the balance
heuristic~\cite{veach1995optimally} to combine sampling of the base surface and
the weighted scratch profiles inside the coherence area into a single unified
sampling strategy.
\paragraph{Combining other BRDFs with our model}
It is of great importance to be able to combine different BRDFs to achieve generality. To this end, we developed a simple modified alpha-blending step that marries correct 
wave-optical shading and mutual interference of scratch contributions to (in principle) arbitrary base BRDFs. In our implementation, we use microfacet models to enable 
a rough base appearance even in unscratched regions. 
{
To achieve this goal, we first retrieve all the scratches from our BVH that fall into the coherence area as before. Next, we calculate the 
weighted scratch area density, i.e. the normalized sum of all contained scratch areas weighted using the Gaussian spatial filter. 
}
This yields a spatially varying ratio between the base and scratch contribution. We use this ratio to blend between the chosen base BRDF and our scratch 
BRDF (\Eq~\ref{eq:scratch_general_brdf}), setting the base amplitude $A'_\base = 0$ in \Eq~\ref{eq:base-response} and $A'_\mask = A'_\scratch = 1$ to ensure energy conservation. 

%
\section{Results}
\label{sec:results}
In this section we will first show example scenes rendered with our model to recreate the appearance of scratched surfaces, 
{
the corresponding render times and parameters can be found in the suplemental material.
}
We will then discuss different aspects of our model in detail, including the impact of the coherent superposition of diffracted light, the possibilities 
to utilize and adapt our model to recreate realistic renderings and finally we will extend our model to incorporate not our specular base response function 
but an arbitrary microfacet model such as GGX~\cite{Trowbridge1975}.
\paragraph{Scratching arbitrary objects}
To facilitate authoring of assets our editing tool allows the alteration of scratch particles in real time by drawing their spatial parameters 
from distributions or applying regular brush drawing techniques in 3D in combination with distribution based alterations of the optical (i.e. profile) parameters. 
\nonarXiv{We provide a detailed video that showcases this tool in the supplementary material of this paper. }Additionally, we have implemented tools that allow conversion of 
2D vector graphics images to scratch particles by projecting them from texture to object space. 
{
In the editor, an approximate real-time BRDF model only shades the first intersected scratch with a single light sample; no coherent effects are taken into account.
}
\paragraph{Coherent vs. incoherent superposition}
The treatment of coherence is of great importance for effects such as mutual interference from structured surfaces, for example compact 
discs or holographic papers. Our model treats spatial coherence by applying a Gaussian weight to the contributions of scratches 
according to their position, as the complex diffracted amplitudes per scratch are superposed. Figure~\ref{fig:incoherent-vs-coherent} reveals that without coherent superposition, effects such as diffraction orders generated by gratings 
are neglected and therefore the associated separation of colors cannot be reproduced. Our model on the other hand is able to closely reproduce such 
phenomena.
\begin{figure}[]%
\centering
 \includegraphics[width=.49\linewidth]{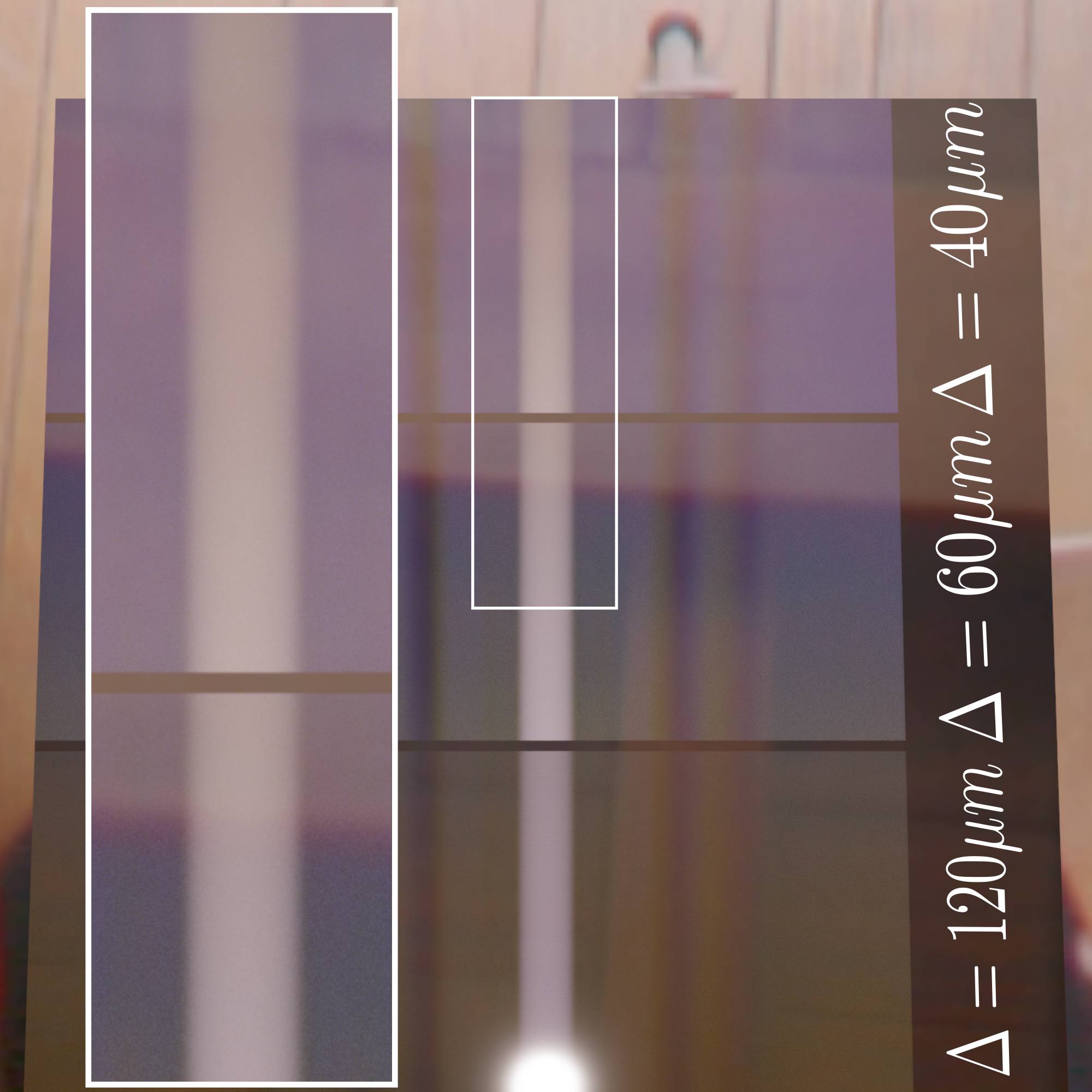}\hfill
 \includegraphics[width=.49\linewidth]{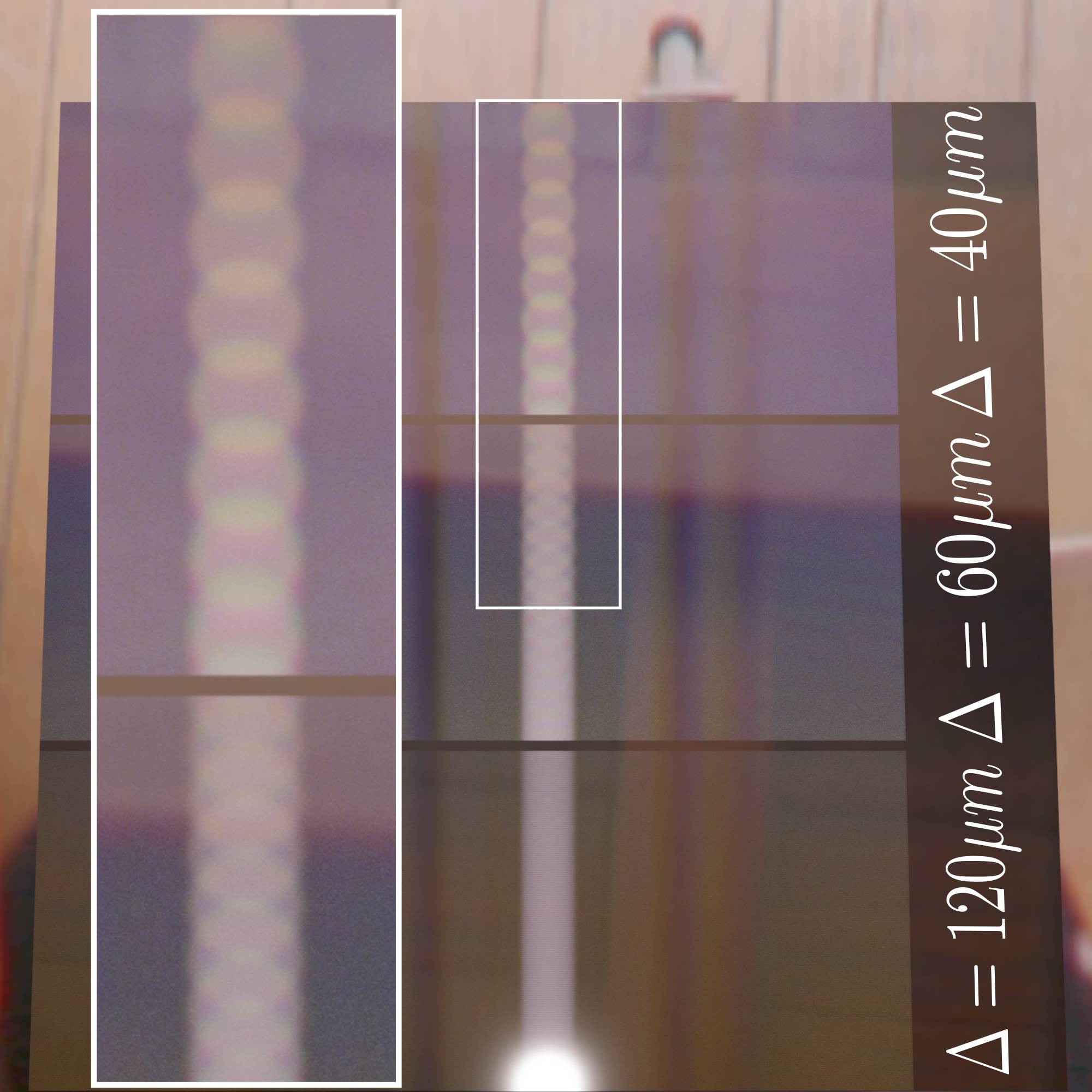}\hfill
\vspace{-2mm}
    \caption{Comparison of coherent and incoherent scratch diffraction superposition. The scratches lie on a metallic plate (GGX microfacet BRDF) and are 
    {
    horizontally arranged as three gratings with different separations $\Delta$ between the uniform scratches. Thus, we expect the diffraction orders only to be visible in vertical direction (across the scratches). 
    Left: An incoherent superposition of scratches within the coherence area leads to colored scratches due to single-scratch diffraction.
    Right: A coherent superposition of scratches (our model) not only accounts for single-scratch diffraction but is also able to recreate mutual interference effects such 
    as higher diffraction orders of the underlying scratch grating which reveals the separation of colors especially in the area of high intensity (see zoom-ins). 
    }
    }
\label{fig:incoherent-vs-coherent}%
\end{figure}
\paragraph{Profile variation}
Scratches on surfaces are created by multiple effects such as every-day wear or even manufacturing. Whereas manufactured scratches or structures 
mostly have a well defined geometry, scratches produced by wear do not. To account for this and more closely reproduce such surfaces, we 
vary width and depth of scratch profiles by sampling from a simplex-noise function~\cite{Perlin2002}. The random number generator used to 
generate the noise is seeded by the scratch index $k$ to ensure determinism. This feature increases realism with very modest impact on performance and memory footprint, 
since longer scratch segments do not need to be split up to incorporate such variations. Figure~\ref{fig:variation} compares the effect of this parameter variation to scratches 
of constant parameters.
\begin{figure}[]%
\centering
\includegraphics[width=0.49\linewidth]{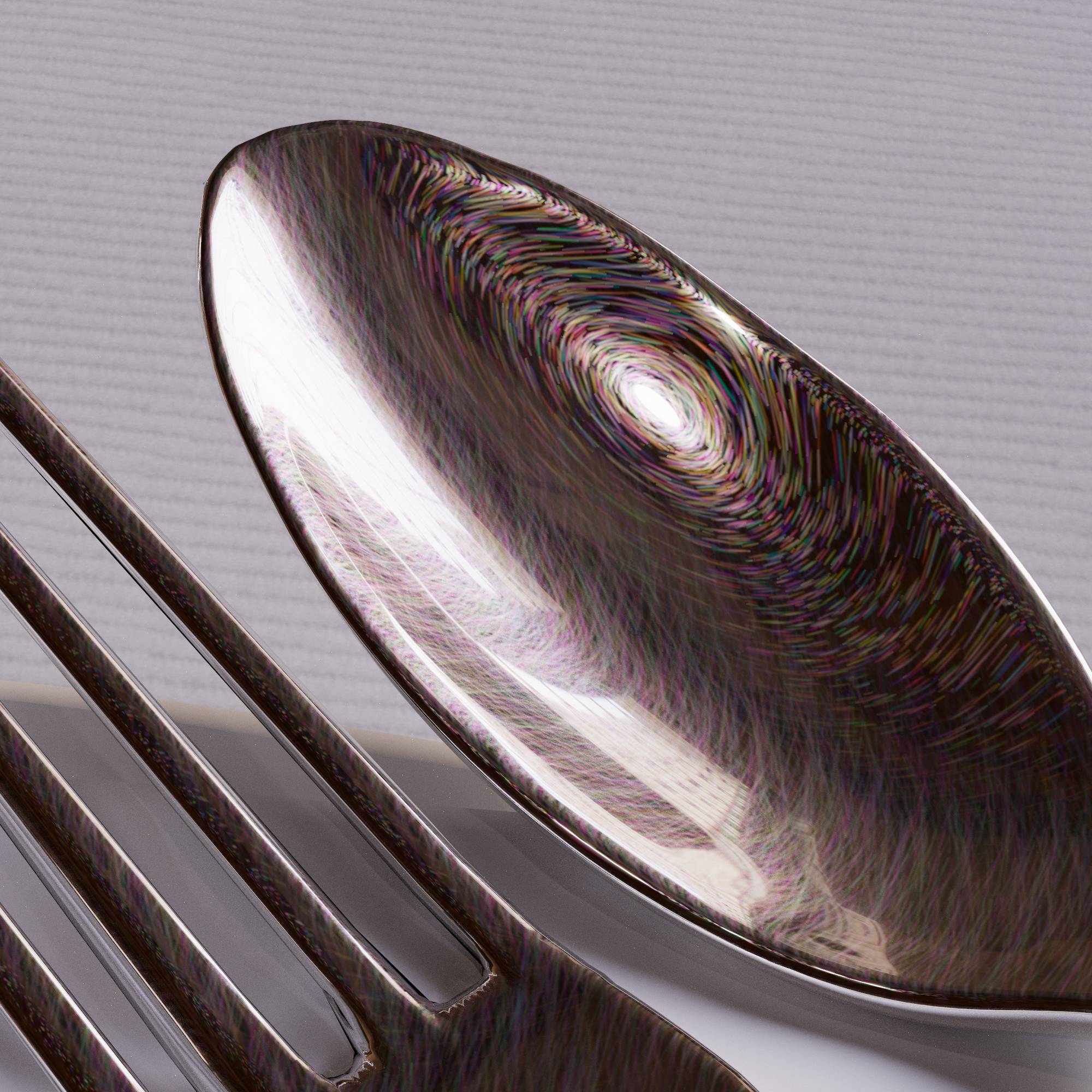}\hfill%
\includegraphics[width=0.49\linewidth]{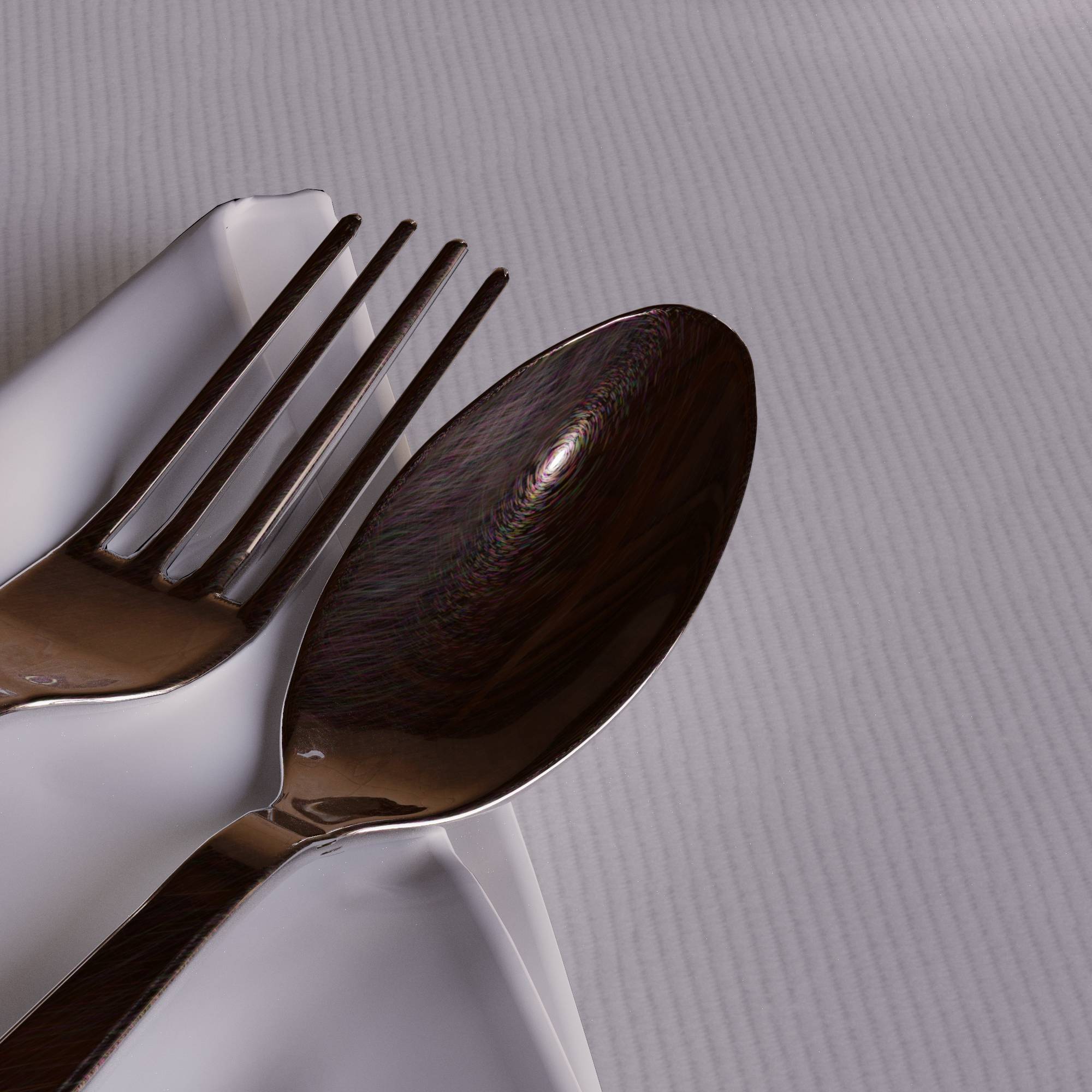}\\
\includegraphics[width=0.49\linewidth]{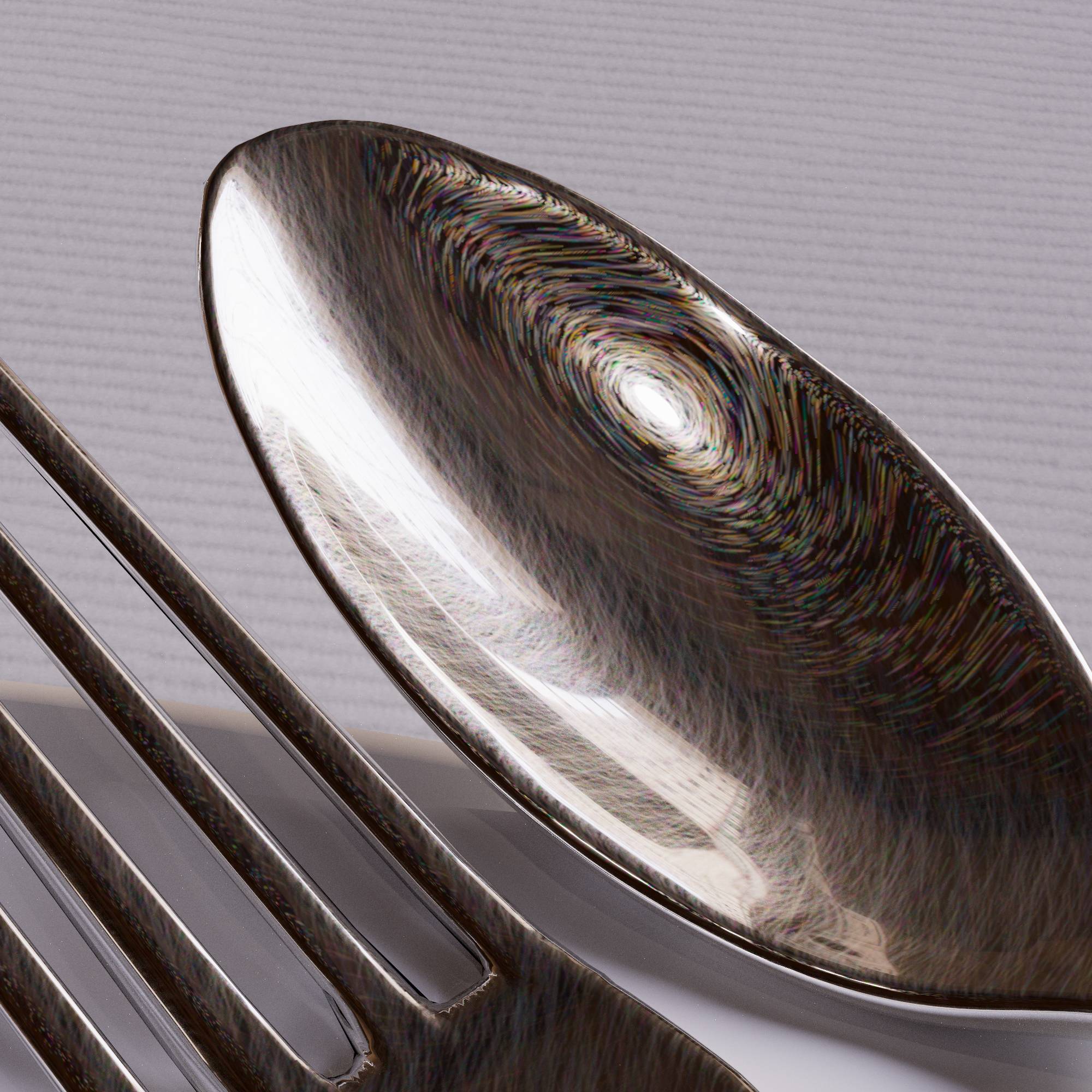}\hfill%
\includegraphics[width=0.49\linewidth]{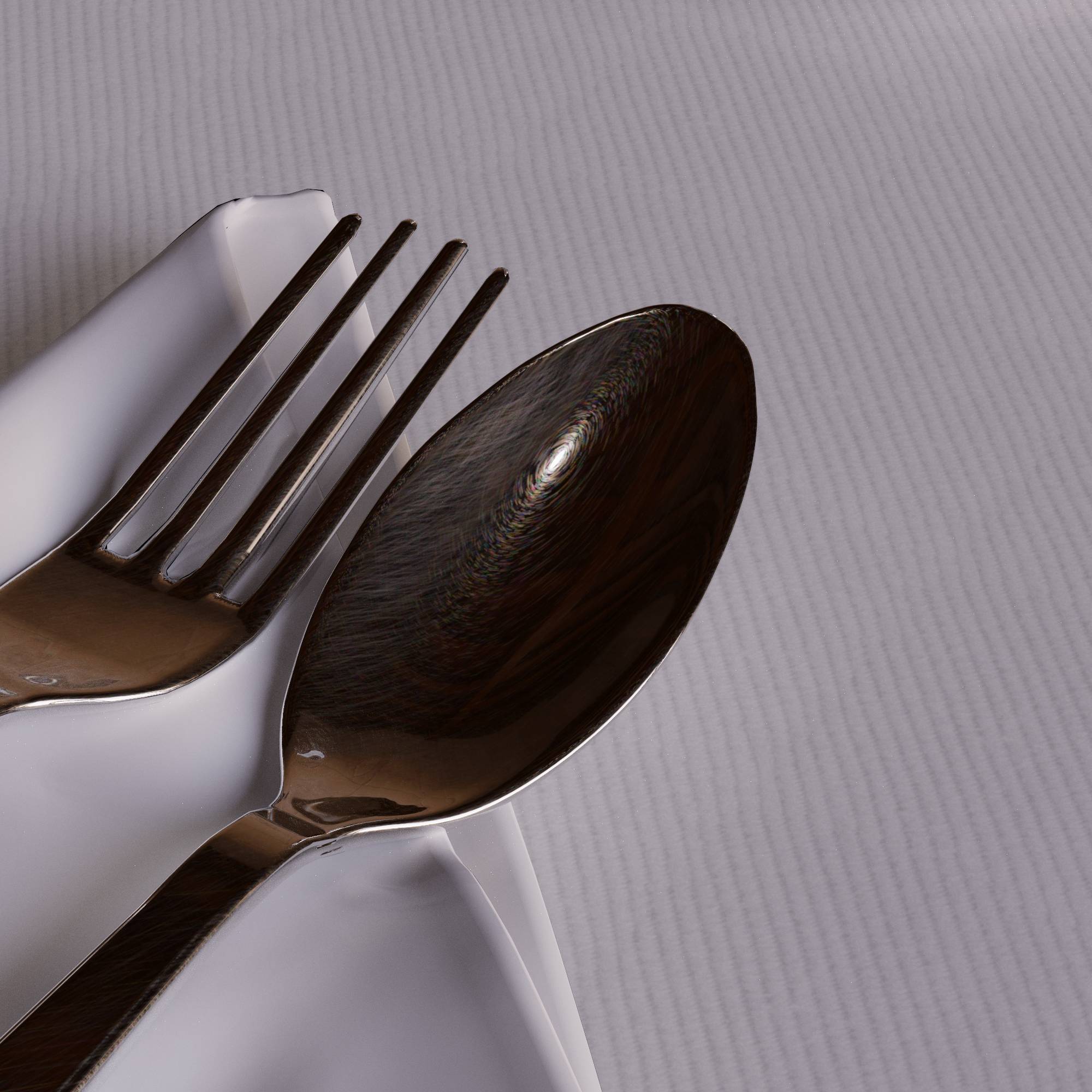}
\caption{Variation of scratch parameters (Top: Without variation) greatly enhances the realistic appearance of our renderings (Bottom: With variation along the scratches)}%
\label{fig:variation}%
\end{figure}
{
\paragraph{Microfacet base blending}
In \Fig~\ref{fig:blending} we show results for our model with our simple blending approach in comparison to the coherent base response. Our fully coherent solution (left) matches well with the specular GGX base (middle) in terms of scratch colors and base reflectance. However, 
some changes can be noticed. First, the specular highlight of our model exhibits a red outline which comes from the fact, that we treat wavelengths separately and thus red light is scattered out compared to smaller wavelengths, an effect which is not covered by geometrical optics models. Second, 
a shift in the color of the scratches can be perceived, which results from the lack of interference with the base. The use of different microfacet models allows us to incorporate surface roughness (right) which, by construction, does not affect the scratches but only the 
base response. In this way we are able to retain the iridescent effects of scratched surfaces with only minor differences and utilize the benefits of microfacet models.
}
\begin{figure*}[t]%
\centering
\includegraphics[width=0.29\linewidth]{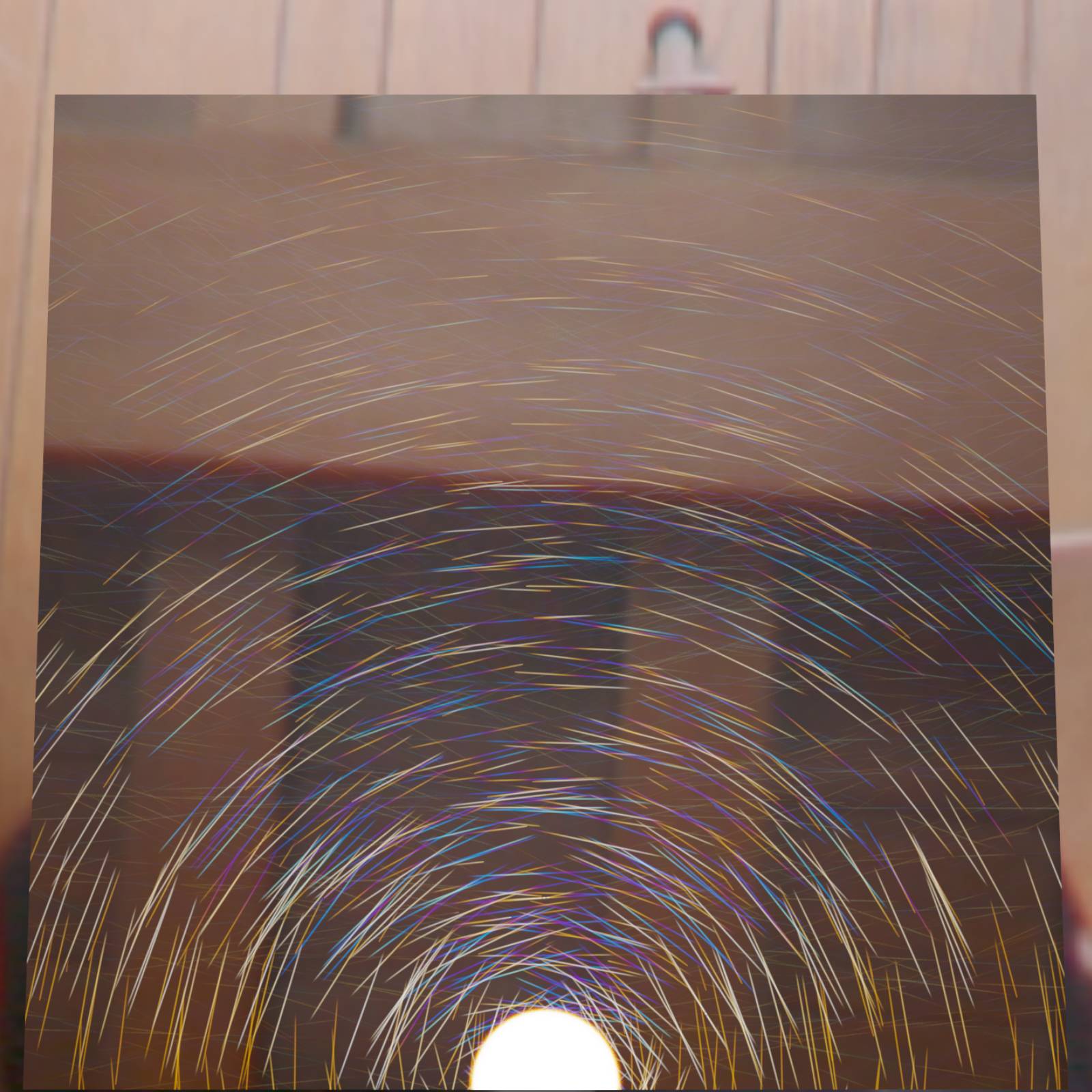}%
\includegraphics[width=0.29\linewidth]{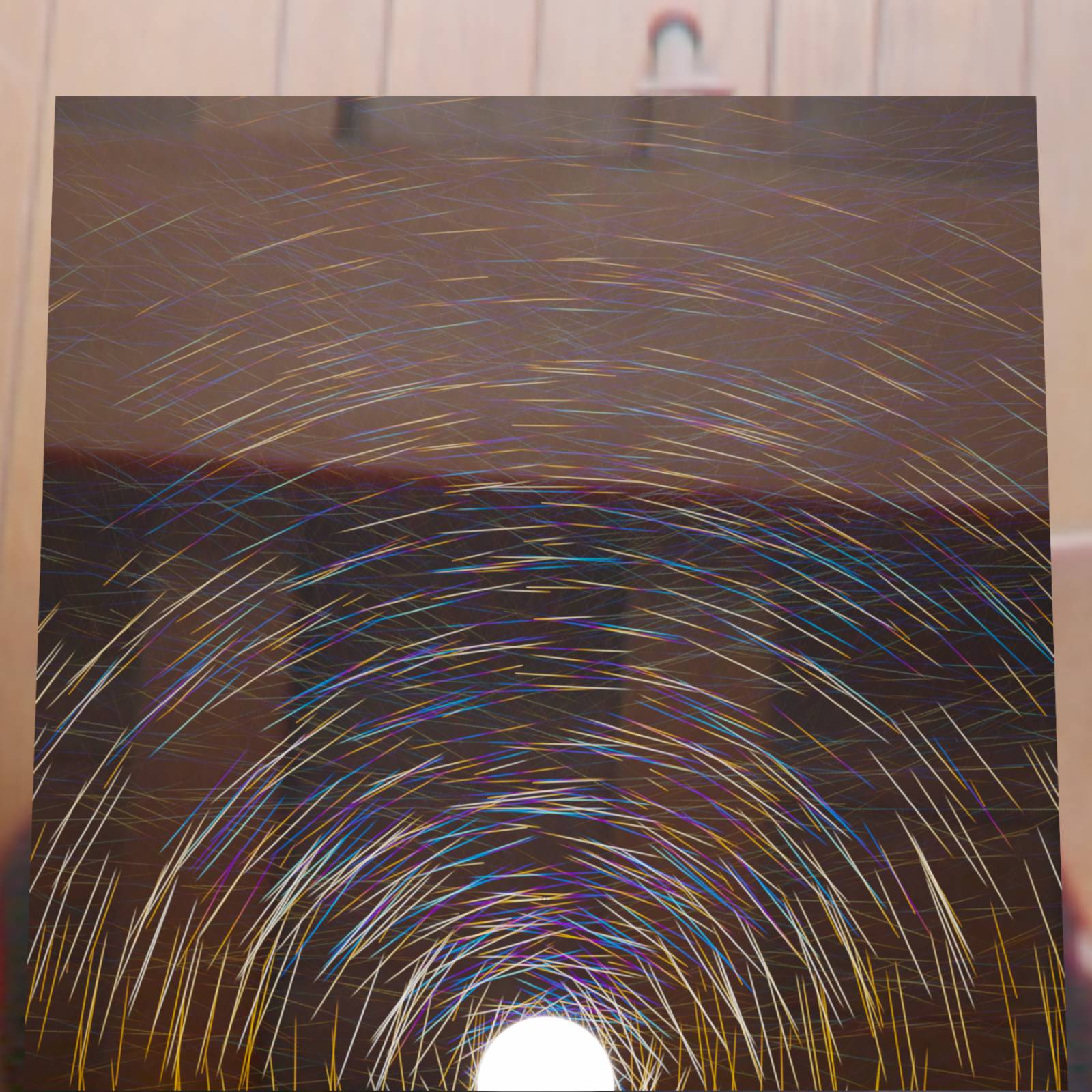}%
\includegraphics[width=0.29\linewidth]{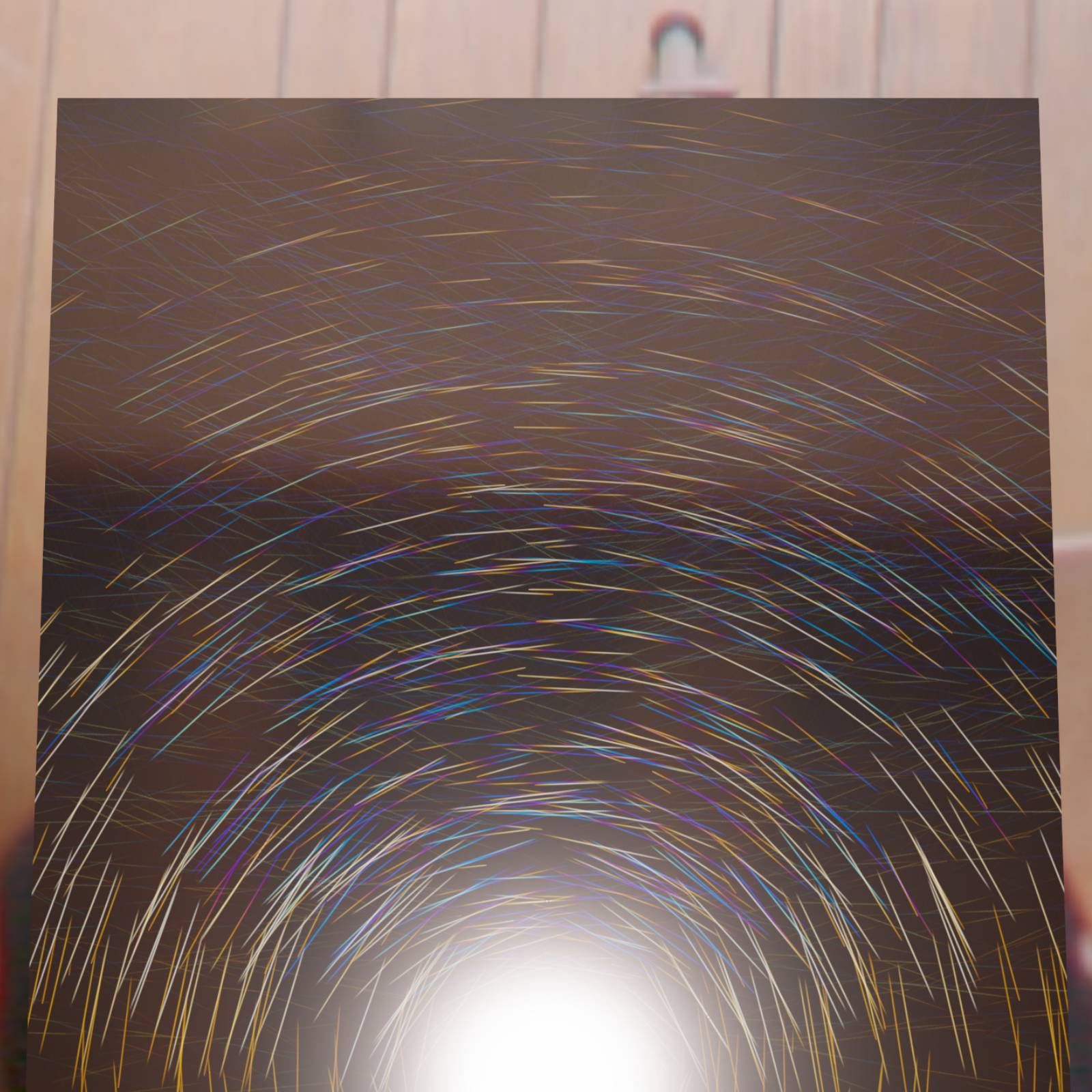}
\caption{Comparison of scratches on a plate with our coherent (left) base response, a smooth/specular GGX (middle) and a rough GGX (right) base reflectance. The appearance of the iridescent colors changes as interference between base and scratches is neglected and single scratch 
diffraction is overestimated}%
\label{fig:blending}%
\end{figure*}

\paragraph{Mapping complex scratch patterns}
Our editing tool also allows us to project (in principle) arbitrarily complex scratch patterns provided as vector graphics onto complex objects. 
\Fig~\ref{fig:pattern-mapping} show the mapping of a vectorized leaf-texture onto a ring.
\begin{figure}[t]%
\centering
\includegraphics[width=0.49\linewidth]{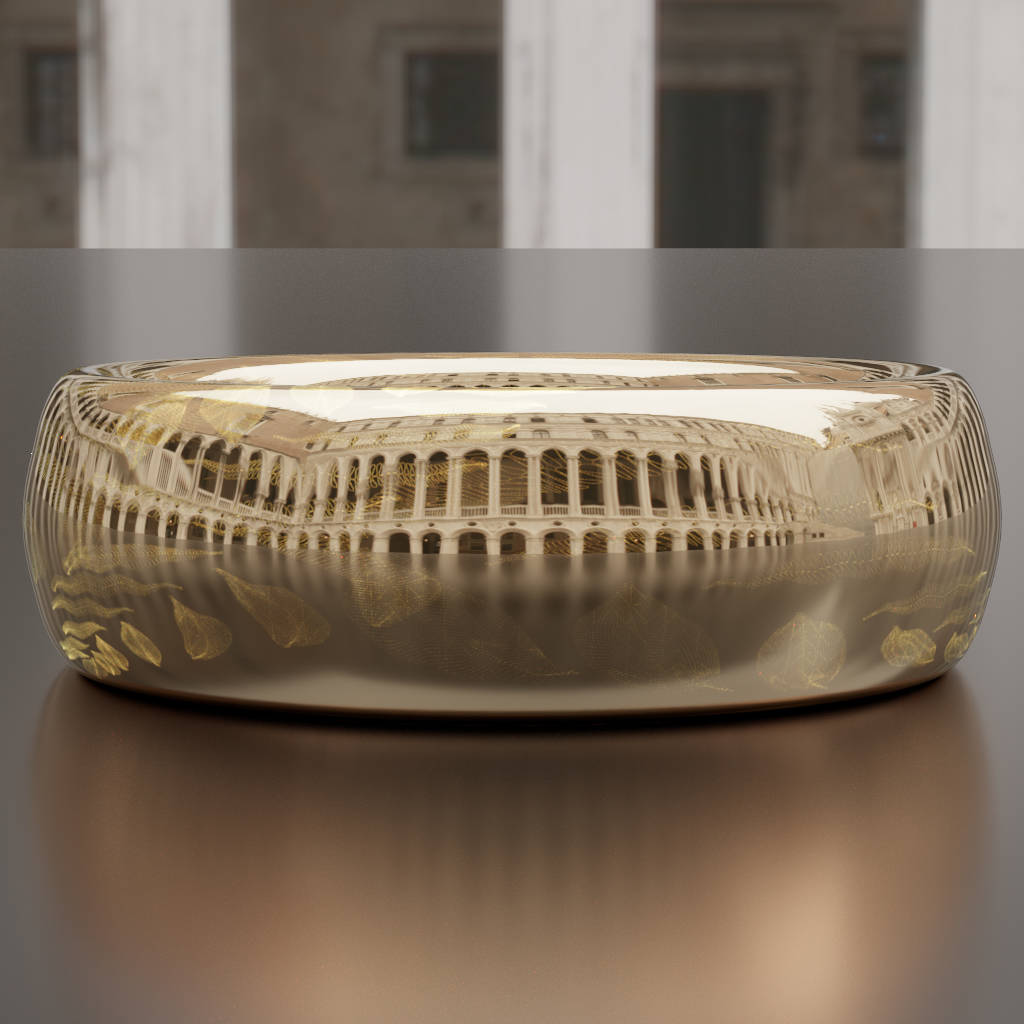}\hfill%
\includegraphics[width=0.49\linewidth]{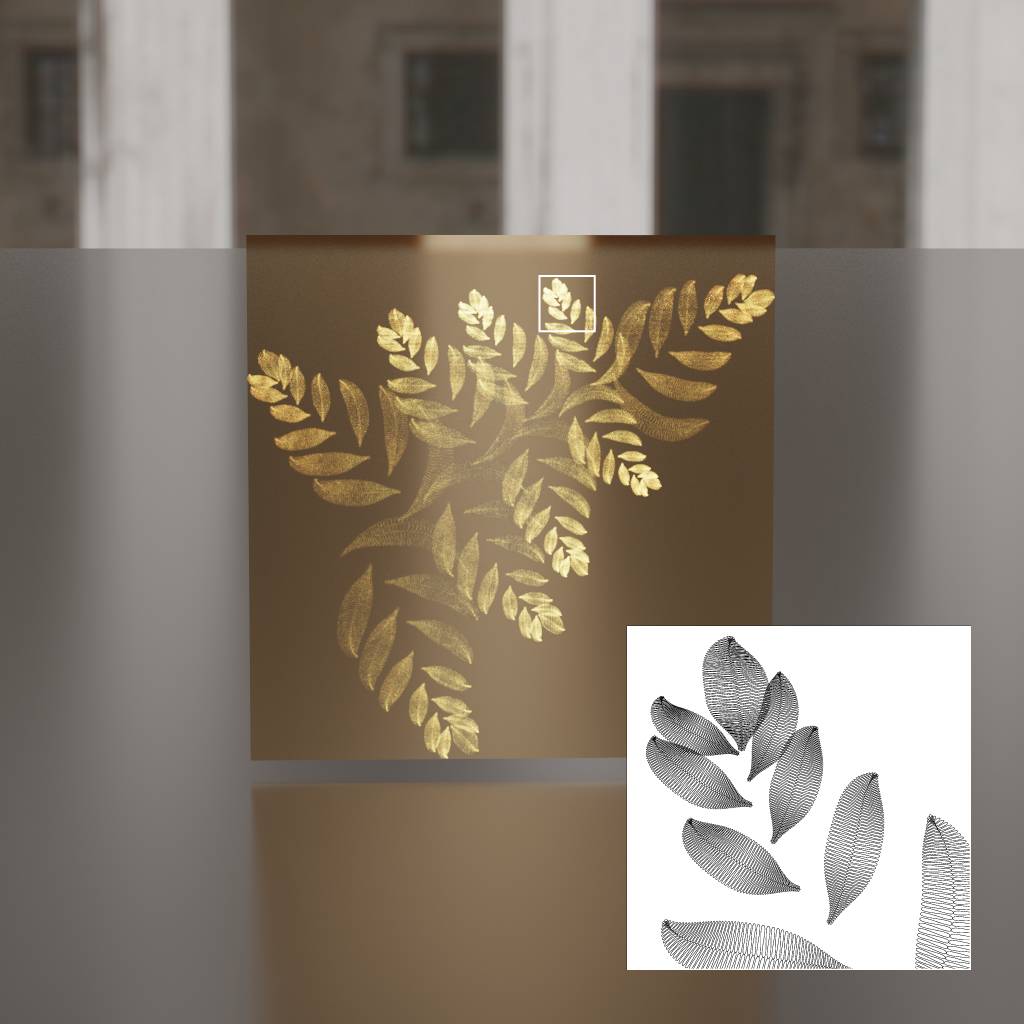}\\
\caption{Our editing tool allows us to map complex scratch patterns onto objects of our choice. Here we ``engraved'' a ring (left) and a planar surface (right) with an intricate vector pattern (inset).}%
\label{fig:pattern-mapping}%
\end{figure}

{
\paragraph{Approximation evaluation}
To evaluate the impact of our assumption of separability (we discard the Gaussian filter in bitangential direction), we compare the numerical radiance obtained via FFT against our model. To this end, we create surfaces which exhibit a number of scratches and 
rasterize these. The resulting heightfield (including scratch intersections) is used to create an optical transfer function via \Eq~\ref{eq:general_otf} and then input into the FFT. The radiance is obtained according to \Eq~\ref{eq:bsdf_coherence} with unit amplitude and the origin as the intersection point. We first show the radiance corresponding to a surface with a single scratch with tangent $(1,0,0)$ (\Fig~\ref{fig:single_scratch_radiance_slices} and \Fig~\ref{fig:scratch_radiance_evaluation}(left)) to clarify single aspects of our approximation and problems that arise when using the FFT. 
A slice along the scratch (left, $\xi'_1$ is the direction cosine in tangential direction) reveals a paraboloid function that rapidly drops off as is expected for a Gaussian in logarithmic representation. We expect a Gaussian as the scratch is longer than the surface we consider, therefore extending the integration limits for $\eta^{(k)}$ to $(-\infty, \infty)$. This is nicely represented by our model and is in good agreement with the numerical solution. 
A slice across the scratch (right, $\xi'_2$ corresponds to the bitangential direction) reveals the approximation of our model. For larger angles we underestimate the radiance which is expected as we neglect the convolution with the Gaussian filter in bitangential direction (c.f. \Eq~\ref{eq:single-scratch-otf-fourier-transform}). This is easily understandable when we consider the convolution as a re-distribution of the diffracted energy, leading to a decrease of the central peak and an increase of the side lobe maxima which is reproduced by the numerical solution. Taking these limitations into account, our model agrees well with the numerical results. 
For a surface with ten randomly distributed scratches (\Fig~\ref{fig:scratch_radiance_evaluation}), the numerical solution for the whole hemisphere (middle) shows severe ghosting artifacts due to discretization which are not present in our analytic model (right), leading to discrepancies between both solutions. However, scales and primary (i.e. non-artifact) radiance distributions agree well, especially along the bitangential directions of the scratches revealing the characteristic diffraction distribution. Differences in tangential directions of scratches result from discretization whereas those in bitangential direction from our approximation as observed for the single scratch case.
\begin{figure*}[t]%
  \includegraphics[width=0.32\linewidth]{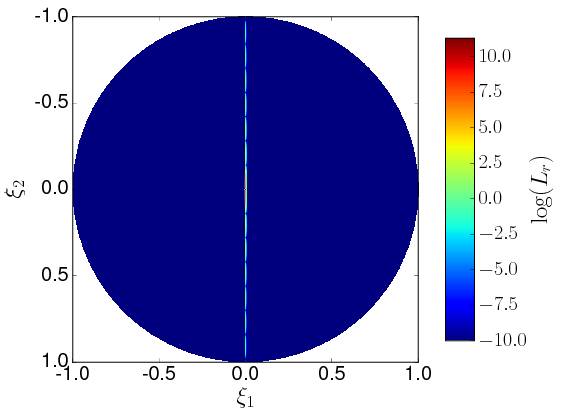}
  \includegraphics[width=0.32\linewidth]{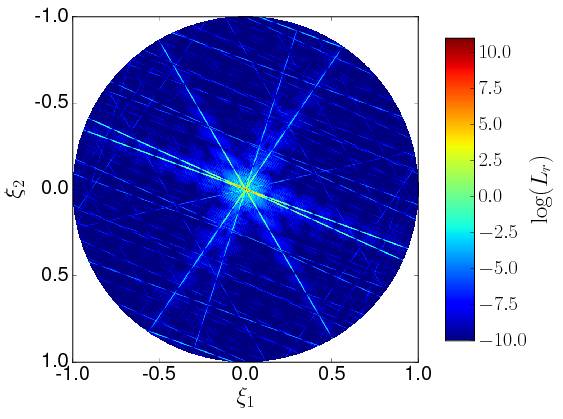}
  \includegraphics[width=0.32\linewidth]{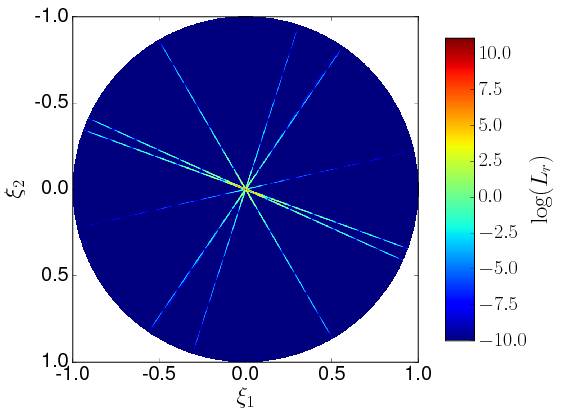}\hfill%
\caption{Numerical radiance obtained via FFT vs. our Model in direction cosine space. {Left}: Analytic single scratch radiance in direction cosine space. 
{Middle; right}: A surface with 10 randomly distributed scratches. The numerical solution for the whole hemisphere (middle) shows severe ghosting artifacts due to discretization which are not present in our analytic model (right). 
\nonarXiv{All plots are available in the supplemental material in high resolution.}}%
\label{fig:scratch_radiance_evaluation}%
\end{figure*}
\begin{figure}
  \includegraphics[width=0.49\linewidth]{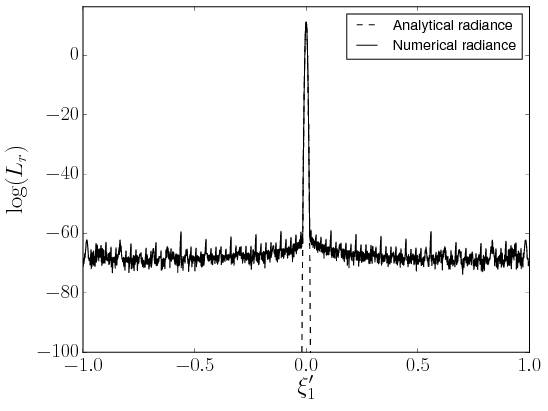}
  \includegraphics[width=0.49\linewidth]{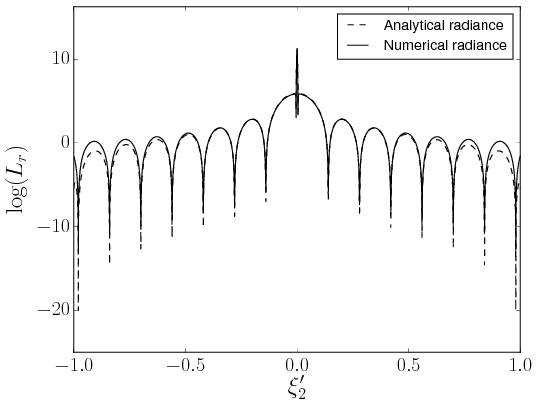}\hfill
  \caption{Comparison of numerical and analytical slices of the radiance distribution for a single scratch (\Fig~\ref{fig:scratch_radiance_evaluation} left). {Left}: $\cvec{\xi}'_2 = 0$: The slice in tangential direction of the scratch reveals the impact of the Gaussian filter. {Right}: $\cvec{\xi}'_1 = 0$: The radiance distribution in bitangential direction reveals diffraction.}%
\label{fig:single_scratch_radiance_slices}%
\end{figure}
}
%
\section{Discussion}
\label{sec:discussion}
In this paper we presented a wave-optical BRDF model for shading surfaces that exhibit microscale scratches. Such scratches 
show iridescent colors under different viewing and lighting conditions resulting from diffraction and mutual interference. 
Our model closely reproduces these effects by encapsulating the wave-optical computations in the shading computation. By 
approximating the spatial coherence using a Gaussian filter we are not only able to recreate localized glint-like 
iridescence but also higher orders of diffraction from grating-like structures. Our approach is very flexible on both the model  and 
the data side. By subdividing paths on the surface into line segments and analytically calculating the corresponding diffracted complex amplitude 
we are able to incorporate arbitrary scratch shapes. The separation of the spatial (paths) and optical (profiles) components of the scratches allows us, in addition, to easily drive and exchange parameters and include variations along the scratch without loss of performance. On the other hand, our editing tool allows us to 
freely scratch arbitrary geometries and apply complex scratching patterns with ease. Still, some parts of the model are really suited for 
improvements in future work:
{
\paragraph{Footprint integration}
Our model recreates local wave-optical diffraction and interference which rely on the surface structures in the vicinity of the shaded point. We therefore subsample the pixel footprint with a spatial filter resembling the coherence function. For full convergence, we rely on 
Monte-Carlo integration performed by the ray-tracer. In future work, an important step to improve performance would be to approximate the integral of the coherent SVBRDF over the pixel footprint to create a full multi-scale model.
}
\paragraph{Importance sampling}
While our base response function is perfectly sampled, and importance sampling strategies for replacement microfacet models readily exist, the sampling strategy 
for the scratch response function uses a uniform sampling for the azimuthal direction, like in d'Eon et al.'s hair model \shortcite{deon2011energy}. 
The azimuthal scattering distribution for scratches is, in fact, quite spiky (according to \Eq~\ref{eq:ex-scratch-response}, 
it is $\rm{sinc}$-distributed for the square profile), so this strategy is overly conservative. 
{In future work, more efficient sampling schemes for typical diffraction patterns and escpecially mutual interference of scratch ensembles (which we consider the problematic part) could be developed.}
\paragraph{Other scratch profiles}
Up to this point we only presented results for a rectangular profile function, showing that it is expressive enough for recreating such intricate effects as iridescent scratches. By incorporating different scratch profiles, we expect to obtain a wider variety of different scattering distributions, and hence an increased degree of realism. Such profiles are not limited to shapes where analytical solutions exist (like V-shaped or Gaussian grooves); in principle, tabulated scattering distributions for arbitrary scratch profiles could be precomputed and used at moderate additional cost.
\paragraph{BSDF extension}
So far, we showed that our model is able to represent the reflection properties of scratched surfaces. The shift invariance of our BRDF on the other hand tells us, 
that the resulting radiance is only dependent on the difference between the incident and outgoing directional cosines. This implies that the model in principle generalizes to transmittance effects (thin materials without internal scattering, like foils or very light curtains) via a simple sign change. This would be an interesting avenue to explore, as many real world diffraction effects are observed in transmittance rather than reflectance.
\paragraph{Coherent base - Use NDF to generate base height variation}
We were able to show that in principle arbitrary BRDFs can be used as a base material for our model based on a modified alpha-blending scheme. 
This approach, however, does not account for phase variations that are directly produced by the material without taking scratches into account. A very interesting 
future direction would be to derive a modified \emph{coherent} base response function that is driven by surface roughness, for instance in the form of microfacet models.
{
\paragraph{Fitting}
Our model consists of (a) light transport simulation and (b) the scene/surface that serves as its input. The first is a physically based model relying on first principles and simplifications necessary to create a feasible approximation. The latter on the other hand 
relates to its own field of research as it can never be fully known except for extremely controlled scenarios like photolithographic surface structures. The full (ground-truth) representation of a real-world surface would require more data than can be practically measured and handled in a rendering context. 
Here, we for the first time introduce a model capable of rendering scratch iridescence that phenomenologically recreates the corresponding effects, namely diffraction and mutual interference. As a next step to test the models' capabilities of approximating the real world, a common approach is to perform fits against real data which, however, requires significant additional research to solve many of the related problems (data acquisition, especially regarding distributions; surface representation) within finite time. We look forward to study these aspects in more detail in future work.
}

\let\OLDthebibliography\thebibliography
\renewcommand\thebibliography[1]{
  \OLDthebibliography{#1}
  \setlength{\parskip}{0pt}
  \setlength{\itemsep}{1.5mm plus 0.3ex}
}
\bibliographystyle{ACM-Reference-Format}

\bibliography{ms}


\begin{thebibliography}{00}


\ifx \showCODEN    \undefined \def \showCODEN     #1{\unskip}     \fi
\ifx \showDOI      \undefined \def \showDOI       #1{#1}\fi
\ifx \showISBNx    \undefined \def \showISBNx     #1{\unskip}     \fi
\ifx \showISBNxiii \undefined \def \showISBNxiii  #1{\unskip}     \fi
\ifx \showISSN     \undefined \def \showISSN      #1{\unskip}     \fi
\ifx \showLCCN     \undefined \def \showLCCN      #1{\unskip}     \fi
\ifx \shownote     \undefined \def \shownote      #1{#1}          \fi
\ifx \showarticletitle \undefined \def \showarticletitle #1{#1}   \fi
\ifx \showURL      \undefined \def \showURL       {\relax}        \fi
\providecommand\bibfield[2]{#2}
\providecommand\bibinfo[2]{#2}
\providecommand\natexlab[1]{#1}
\providecommand\showeprint[2][]{arXiv:#2}

\bibitem[\protect\citeauthoryear{Bosch, Pueyo, Merillou, and
  Ghazanfarpour}{Bosch et~al\mbox{.}}{2008}]%
        {Bosch2008}
\bibfield{author}{\bibinfo{person}{C. Bosch}, \bibinfo{person}{X. Pueyo},
  \bibinfo{person}{S. Merillou}, {and} \bibinfo{person}{D. Ghazanfarpour}.}
  \bibinfo{year}{2008}\natexlab{}.
\newblock \showarticletitle{{A Resolution Independent Approach for the Accurate
  Rendering of Grooved Surfaces}}.
\newblock \bibinfo{journal}{{\em Computer Graphics Forum\/}}
  (\bibinfo{year}{2008}).
\newblock
\showISSN{1467-8659}
\showDOI{%
\url{https://doi.org/10.1111/j.1467-8659.2008.01342.x}}


\bibitem[\protect\citeauthoryear{Bosch, Pueyo, Mérillou, and
  Ghazanfarpour}{Bosch et~al\mbox{.}}{2004}]%
        {Bosch2004}
\bibfield{author}{\bibinfo{person}{Carles Bosch}, \bibinfo{person}{Xavier
  Pueyo}, \bibinfo{person}{Stéphane Mérillou}, {and}
  \bibinfo{person}{Djamchid Ghazanfarpour}.} \bibinfo{year}{2004}\natexlab{}.
\newblock \showarticletitle{A Physically-Based Model for Rendering Realistic
  Scratches}.
\newblock \bibinfo{journal}{{\em Computer Graphics Forum\/}}
  \bibinfo{volume}{23}, \bibinfo{number}{3} (\bibinfo{year}{2004}),
  \bibinfo{pages}{361--370}.
\newblock
\showISSN{1467-8659}
\showDOI{%
\url{https://doi.org/10.1111/j.1467-8659.2004.00767.x}}


\bibitem[\protect\citeauthoryear{Buchanan and Lalonde}{Buchanan and
  Lalonde}{1999}]%
        {Buchanan1999}
\bibfield{author}{\bibinfo{person}{John~W. Buchanan} {and}
  \bibinfo{person}{Paul Lalonde}.} \bibinfo{year}{1999}\natexlab{}.
\newblock \showarticletitle{An Observational Model for Illuminating Isolated
  Scratches.}
\newblock \bibinfo{journal}{{\em Proc. Western Computer Graphics Symposium 1999
  (SKIGRAPH’99)\/}} (\bibinfo{year}{1999}).
\newblock


\bibitem[\protect\citeauthoryear{Church and Takacs}{Church and Takacs}{2009}]%
        {Church1995}
\bibfield{author}{\bibinfo{person}{Eugene~L Church} {and}
  \bibinfo{person}{Peter~Z Takacs}.} \bibinfo{year}{2009}\natexlab{}.
\newblock \showarticletitle{Surface scattering}.
\newblock In \bibinfo{booktitle}{{\em Handbook of Optics}},
  \bibfield{editor}{\bibinfo{person}{{Bass, M.~et al.}}} (Ed.).
  \bibinfo{publisher}{McGraw-Hill}, Chapter~7, \bibinfo{pages}{7.1--7.14}.
\newblock


\bibitem[\protect\citeauthoryear{Cuypers, Haber, Bekaert, Oh, and
  Raskar}{Cuypers et~al\mbox{.}}{2012}]%
        {Cuypers2012}
\bibfield{author}{\bibinfo{person}{Tom Cuypers}, \bibinfo{person}{Tom Haber},
  \bibinfo{person}{Philippe Bekaert}, \bibinfo{person}{Se~Baek Oh}, {and}
  \bibinfo{person}{Ramesh Raskar}.} \bibinfo{year}{2012}\natexlab{}.
\newblock \showarticletitle{Reflectance Model for Diffraction}.
\newblock \bibinfo{journal}{{\em ACM Trans. Graph.\/}} \bibinfo{volume}{31},
  \bibinfo{number}{5}, Article \bibinfo{articleno}{122} (\bibinfo{date}{sep}
  \bibinfo{year}{2012}), \bibinfo{numpages}{11}~pages.
\newblock
\showISSN{0730-0301}
\showDOI{%
\url{https://doi.org/10.1145/2231816.2231820}}


\bibitem[\protect\citeauthoryear{d'Eon, Francois, Hill, Letteri, and
  Aubry}{d'Eon et~al\mbox{.}}{2011}]%
        {deon2011energy}
\bibfield{author}{\bibinfo{person}{Eugene d'Eon}, \bibinfo{person}{Guillaume
  Francois}, \bibinfo{person}{Martin Hill}, \bibinfo{person}{Joe Letteri},
  {and} \bibinfo{person}{Jean-Marie Aubry}.} \bibinfo{year}{2011}\natexlab{}.
\newblock \showarticletitle{An Energy-Conserving Hair Reflectance Model}.
\newblock \bibinfo{journal}{{\em Computer Graphics Forum\/}}
  \bibinfo{volume}{30}, \bibinfo{number}{4} (\bibinfo{year}{2011}).
\newblock
\showISSN{1467-8659}


\bibitem[\protect\citeauthoryear{Dhillon, Teyssier, Single, Gaponenko,
  Milinkovitch, and Zwicker}{Dhillon et~al\mbox{.}}{2014}]%
        {Dhillon2014}
\bibfield{author}{\bibinfo{person}{D.S. Dhillon}, \bibinfo{person}{J.
  Teyssier}, \bibinfo{person}{M. Single}, \bibinfo{person}{I. Gaponenko},
  \bibinfo{person}{M.C. Milinkovitch}, {and} \bibinfo{person}{M. Zwicker}.}
  \bibinfo{year}{2014}\natexlab{}.
\newblock \showarticletitle{Interactive Diffraction from Biological
  Nanostructures}.
\newblock \bibinfo{journal}{{\em Comput. Graph. Forum\/}} \bibinfo{volume}{33},
  \bibinfo{number}{8} (\bibinfo{date}{dec} \bibinfo{year}{2014}),
  \bibinfo{pages}{177--188}.
\newblock
\showISSN{0167-7055}
\showDOI{%
\url{https://doi.org/10.1111/cgf.12425}}


\bibitem[\protect\citeauthoryear{Divitt and Novotny}{Divitt and
  Novotny}{2015}]%
        {Divitt2015}
\bibfield{author}{\bibinfo{person}{Shawn Divitt} {and} \bibinfo{person}{Lukas
  Novotny}.} \bibinfo{year}{2015}\natexlab{}.
\newblock \showarticletitle{Spatial coherence of sunlight and its implications
  for light management in photovoltaics}.
\newblock \bibinfo{journal}{{\em Optica\/}} \bibinfo{volume}{2},
  \bibinfo{number}{2} (\bibinfo{date}{Feb} \bibinfo{year}{2015}),
  \bibinfo{pages}{95--103}.
\newblock
\showDOI{%
\url{https://doi.org/10.1364/OPTICA.2.000095}}


\bibitem[\protect\citeauthoryear{Dong, Wang, Tong, Snyder, Lan, Ben-Ezra, and
  Guo}{Dong et~al\mbox{.}}{2010}]%
        {Yue2010}
\bibfield{author}{\bibinfo{person}{Yue Dong}, \bibinfo{person}{Jiaping Wang},
  \bibinfo{person}{Xin Tong}, \bibinfo{person}{John Snyder},
  \bibinfo{person}{Yanxiang Lan}, \bibinfo{person}{Moshe Ben-Ezra}, {and}
  \bibinfo{person}{Baining Guo}.} \bibinfo{year}{2010}\natexlab{}.
\newblock \showarticletitle{Manifold Bootstrapping for {SVBRDF} Capture}.
\newblock \bibinfo{journal}{{\em ACM Trans. Graph.\/}} \bibinfo{volume}{29},
  \bibinfo{number}{4} (\bibinfo{year}{2010}), \bibinfo{pages}{98:1--98:10}.
\newblock


\bibitem[\protect\citeauthoryear{Dong, Walter, Marschner, and Greenberg}{Dong
  et~al\mbox{.}}{2015}]%
        {Dong2015}
\bibfield{author}{\bibinfo{person}{Zhao Dong}, \bibinfo{person}{Bruce Walter},
  \bibinfo{person}{Steve Marschner}, {and} \bibinfo{person}{Donald~P.
  Greenberg}.} \bibinfo{year}{2015}\natexlab{}.
\newblock \showarticletitle{Predicting Appearance from Measured Microgeometry
  of Metal Surfaces}.
\newblock \bibinfo{journal}{{\em ACM Trans. Graph.\/}} \bibinfo{volume}{35},
  \bibinfo{number}{1}, Article \bibinfo{articleno}{9} (\bibinfo{date}{Dec.}
  \bibinfo{year}{2015}), \bibinfo{numpages}{13}~pages.
\newblock
\showISSN{0730-0301}
\showDOI{%
\url{https://doi.org/10.1145/2815618}}


\bibitem[\protect\citeauthoryear{Dorsey, Rushmeier, and Sillion}{Dorsey
  et~al\mbox{.}}{2010}]%
        {dorsey2010digital}
\bibfield{author}{\bibinfo{person}{Julie Dorsey}, \bibinfo{person}{Holly
  Rushmeier}, {and} \bibinfo{person}{Fran{\c{c}}ois Sillion}.}
  \bibinfo{year}{2010}\natexlab{}.
\newblock \bibinfo{booktitle}{{\em Digital modeling of material appearance}}.
\newblock \bibinfo{publisher}{Morgan Kaufmann}.
\newblock


\bibitem[\protect\citeauthoryear{Goodman}{Goodman}{1996}]%
        {Goodman1996}
\bibfield{author}{\bibinfo{person}{J.W. Goodman}.}
  \bibinfo{year}{1996}\natexlab{}.
\newblock \bibinfo{booktitle}{{\em Introduction to Fourier Optics}}.
\newblock \bibinfo{publisher}{McGraw-Hill}.
\newblock
\showISBNx{9780070242548}
\showLCCN{95082033}


\bibitem[\protect\citeauthoryear{Harvey, Vernold, Krywonos, and
  Thompson}{Harvey et~al\mbox{.}}{2000}]%
        {Harvey2000}
\bibfield{author}{\bibinfo{person}{James~E. Harvey},
  \bibinfo{person}{Cynthia~L. Vernold}, \bibinfo{person}{Andrey Krywonos},
  {and} \bibinfo{person}{Patrick~L. Thompson}.}
  \bibinfo{year}{2000}\natexlab{}.
\newblock \showarticletitle{Diffracted radiance: a fundamental quantity in
  nonparaxial scalar diffraction theory: errata}.
\newblock \bibinfo{journal}{{\em Appl. Opt.\/}} \bibinfo{volume}{39},
  \bibinfo{number}{34} (\bibinfo{date}{Dec} \bibinfo{year}{2000}),
  \bibinfo{pages}{6374--6375}.
\newblock
\showDOI{%
\url{https://doi.org/10.1364/AO.39.006374}}


\bibitem[\protect\citeauthoryear{He, Torrance, Sillion, and Greenberg}{He
  et~al\mbox{.}}{1991}]%
        {He1991}
\bibfield{author}{\bibinfo{person}{Xiao~D. He}, \bibinfo{person}{Kenneth~E.
  Torrance}, \bibinfo{person}{Fran\c{c}ois~X. Sillion}, {and}
  \bibinfo{person}{Donald~P. Greenberg}.} \bibinfo{year}{1991}\natexlab{}.
\newblock \showarticletitle{A Comprehensive Physical Model for Light
  Reflection}. In \bibinfo{booktitle}{{\em Proceedings of the 18th Annual
  Conference on Computer Graphics and Interactive Techniques}} {\em
  (\bibinfo{series}{SIGGRAPH '91})}. \bibinfo{publisher}{ACM},
  \bibinfo{address}{New York, NY, USA}, \bibinfo{pages}{175--186}.
\newblock
\showISBNx{0-89791-436-8}
\showDOI{%
\url{https://doi.org/10.1145/122718.122738}}


\bibitem[\protect\citeauthoryear{Holzschuch and Pacanowski}{Holzschuch and
  Pacanowski}{2016}]%
        {Holzschuch2016}
\bibfield{author}{\bibinfo{person}{Nicolas Holzschuch} {and}
  \bibinfo{person}{Romain Pacanowski}.} \bibinfo{year}{2016}\natexlab{}.
\newblock \bibinfo{booktitle}{{\em {A Physically-Based Reflectance Model
  Combining Reflection and Diffraction}}}.
\newblock \bibinfo{type}{Research Report} RR-8964.
  \bibinfo{institution}{{INRIA}}.
\newblock
\showURL{%
\url{https://hal.inria.fr/hal-01386157}}


\bibitem[\protect\citeauthoryear{Hullin, Eisemann, Seidel, and Lee}{Hullin
  et~al\mbox{.}}{2011}]%
        {HullinSIG2011}
\bibfield{author}{\bibinfo{person}{Matthias~B. Hullin}, \bibinfo{person}{Elmar
  Eisemann}, \bibinfo{person}{Hans-Peter Seidel}, {and}
  \bibinfo{person}{Sungkil Lee}.} \bibinfo{year}{2011}\natexlab{}.
\newblock \showarticletitle{Physically-Based Real-Time Lens Flare Rendering}.
\newblock \bibinfo{journal}{{\em ACM Trans. Graph. (Proc. SIGGRAPH 2011)\/}}
  \bibinfo{volume}{30}, \bibinfo{number}{4} (\bibinfo{year}{2011}),
  \bibinfo{pages}{108:1--108:9}.
\newblock


\bibitem[\protect\citeauthoryear{Krywonos}{Krywonos}{2006}]%
        {Krywonos2006}
\bibfield{author}{\bibinfo{person}{Krywonos}.} \bibinfo{year}{2006}\natexlab{}.
\newblock \bibinfo{booktitle}{{\em Predicting Surface Scatter Using a Linear
  Systems Formulation of Non-paraxial Scalar Diffraction}}.
\newblock \bibinfo{publisher}{University of Central Florida}.
\newblock
\showISBNx{9780542975868}
\showURL{%
\url{http://etd.fcla.edu/CF/CFE0001446/Krywonos_Andrey_200612_PhD.pdf}}


\bibitem[\protect\citeauthoryear{Lauterbach, Garland, Sengupta, Luebke, and
  Manocha}{Lauterbach et~al\mbox{.}}{2009}]%
        {Lauterbach2009}
\bibfield{author}{\bibinfo{person}{C. Lauterbach}, \bibinfo{person}{M.
  Garland}, \bibinfo{person}{S. Sengupta}, \bibinfo{person}{D. Luebke}, {and}
  \bibinfo{person}{D. Manocha}.} \bibinfo{year}{2009}\natexlab{}.
\newblock \showarticletitle{{Fast BVH Construction on GPUs}}.
\newblock \bibinfo{journal}{{\em Computer Graphics Forum\/}}
  (\bibinfo{year}{2009}).
\newblock
\showISSN{1467-8659}
\showDOI{%
\url{https://doi.org/10.1111/j.1467-8659.2009.01377.x}}


\bibitem[\protect\citeauthoryear{Levin, Glasner, Xiong, Durand, Freeman,
  Matusik, and Zickler}{Levin et~al\mbox{.}}{2013}]%
        {Levin2013}
\bibfield{author}{\bibinfo{person}{Anat Levin}, \bibinfo{person}{Daniel
  Glasner}, \bibinfo{person}{Ying Xiong}, \bibinfo{person}{Fr{\'e}do Durand},
  \bibinfo{person}{William Freeman}, \bibinfo{person}{Wojciech Matusik}, {and}
  \bibinfo{person}{Todd Zickler}.} \bibinfo{year}{2013}\natexlab{}.
\newblock \showarticletitle{Fabricating {BRDF}s at High Spatial Resolution
  Using Wave Optics}.
\newblock \bibinfo{journal}{{\em ACM Trans. Graph.\/}} \bibinfo{volume}{32},
  \bibinfo{number}{4}, Article \bibinfo{articleno}{144} (\bibinfo{date}{July}
  \bibinfo{year}{2013}), \bibinfo{numpages}{14}~pages.
\newblock
\showISSN{0730-0301}
\showDOI{%
\url{https://doi.org/10.1145/2461912.2461981}}


\bibitem[\protect\citeauthoryear{Lipson, Lipson, and Lipson}{Lipson
  et~al\mbox{.}}{2010}]%
        {Lipson2010}
\bibfield{author}{\bibinfo{person}{Ariel Lipson}, \bibinfo{person}{Stephen~G
  Lipson}, {and} \bibinfo{person}{Henry Lipson}.}
  \bibinfo{year}{2010}\natexlab{}.
\newblock \bibinfo{booktitle}{{\em Optical physics}}.
\newblock \bibinfo{publisher}{Cambridge University Press},
  \bibinfo{address}{Leiden}.
\newblock
\showURL{%
\url{https://cds.cern.ch/record/1338386}}


\bibitem[\protect\citeauthoryear{L\"{o}w, Kronander, Ynnerman, and
  Unger}{L\"{o}w et~al\mbox{.}}{2012}]%
        {Loew2012}
\bibfield{author}{\bibinfo{person}{Joakim L\"{o}w}, \bibinfo{person}{Joel
  Kronander}, \bibinfo{person}{Anders Ynnerman}, {and} \bibinfo{person}{Jonas
  Unger}.} \bibinfo{year}{2012}\natexlab{}.
\newblock \showarticletitle{{BRDF} Models for Accurate and Efficient Rendering
  of Glossy Surfaces}.
\newblock \bibinfo{journal}{{\em ACM Trans. Graph.\/}} \bibinfo{volume}{31},
  \bibinfo{number}{1}, Article \bibinfo{articleno}{9} (\bibinfo{date}{Feb.}
  \bibinfo{year}{2012}), \bibinfo{numpages}{14}~pages.
\newblock
\showISSN{0730-0301}
\showDOI{%
\url{https://doi.org/10.1145/2077341.2077350}}


\bibitem[\protect\citeauthoryear{Lu, Koenderink, and Kappers}{Lu
  et~al\mbox{.}}{2000}]%
        {Lu2000}
\bibfield{author}{\bibinfo{person}{Rong Lu}, \bibinfo{person}{Jan~J.
  Koenderink}, {and} \bibinfo{person}{Astrid~M.L. Kappers}.}
  \bibinfo{year}{2000}\natexlab{}.
\newblock \showarticletitle{Specularities on Surfaces with Tangential Hairs or
  Grooves}.
\newblock \bibinfo{journal}{{\em Comput. Vis. Image Underst.\/}}
  \bibinfo{volume}{78}, \bibinfo{number}{3} (\bibinfo{date}{jun}
  \bibinfo{year}{2000}), \bibinfo{pages}{320--335}.
\newblock
\showISSN{1077-3142}
\showDOI{%
\url{https://doi.org/10.1006/cviu.2000.0841}}


\bibitem[\protect\citeauthoryear{Mandel and Wolf}{Mandel and Wolf}{1995}]%
        {MandelWolf1995}
\bibfield{author}{\bibinfo{person}{Leonard Mandel} {and} \bibinfo{person}{Emil
  Wolf}.} \bibinfo{year}{1995}\natexlab{}.
\newblock \bibinfo{booktitle}{{\em Optical Coherence and Quantum Optics}}.
\newblock \bibinfo{publisher}{Cambridge University Press}.
\newblock
\showISBNx{9780521417112}


\bibitem[\protect\citeauthoryear{Marschner, Westin, Arbree, and Moon}{Marschner
  et~al\mbox{.}}{2005}]%
        {Marschner2005}
\bibfield{author}{\bibinfo{person}{Stephen~R. Marschner},
  \bibinfo{person}{Stephen~H. Westin}, \bibinfo{person}{Adam Arbree}, {and}
  \bibinfo{person}{Jonathan~T. Moon}.} \bibinfo{year}{2005}\natexlab{}.
\newblock \showarticletitle{Measuring and Modeling the Appearance of Finished
  Wood}.
\newblock \bibinfo{journal}{{\em ACM Trans. Graph.\/}} \bibinfo{volume}{24},
  \bibinfo{number}{3} (\bibinfo{date}{July} \bibinfo{year}{2005}),
  \bibinfo{pages}{727--734}.
\newblock


\bibitem[\protect\citeauthoryear{Merillou, Dischler, and
  Ghazanfarpour}{Merillou et~al\mbox{.}}{2001}]%
        {Merillou2001}
\bibfield{author}{\bibinfo{person}{S. Merillou}, \bibinfo{person}{J.M.
  Dischler}, {and} \bibinfo{person}{D. Ghazanfarpour}.}
  \bibinfo{year}{2001}\natexlab{}.
\newblock \showarticletitle{Surface scratches: measuring, modeling and
  rendering}.
\newblock \bibinfo{journal}{{\em The Visual Computer\/}} \bibinfo{volume}{17},
  \bibinfo{number}{1} (\bibinfo{year}{2001}), \bibinfo{pages}{30--45}.
\newblock
\showISSN{1432-2315}
\showDOI{%
\url{https://doi.org/10.1007/s003710000093}}


\bibitem[\protect\citeauthoryear{Musbach, Meyer, Reitich, and Oh}{Musbach
  et~al\mbox{.}}{2013}]%
        {Musbach2012}
\bibfield{author}{\bibinfo{person}{A. Musbach}, \bibinfo{person}{G.~W. Meyer},
  \bibinfo{person}{F. Reitich}, {and} \bibinfo{person}{S.~H. Oh}.}
  \bibinfo{year}{2013}\natexlab{}.
\newblock \showarticletitle{Full Wave Modelling of Light Propagation and
  Reflection}.
\newblock \bibinfo{journal}{{\em Computer Graphics Forum\/}}
  \bibinfo{volume}{32}, \bibinfo{number}{6} (\bibinfo{year}{2013}),
  \bibinfo{pages}{24--37}.
\newblock
\showISSN{1467-8659}
\showDOI{%
\url{https://doi.org/10.1111/cgf.12012}}


\bibitem[\protect\citeauthoryear{Ngan, Durand, and Matusik}{Ngan
  et~al\mbox{.}}{2005}]%
        {Ngan2005b}
\bibfield{author}{\bibinfo{person}{Addy Ngan}, \bibinfo{person}{Fr{\'e}do
  Durand}, {and} \bibinfo{person}{Wojciech Matusik}.}
  \bibinfo{year}{2005}\natexlab{}.
\newblock \showarticletitle{Experimental Analysis of {BRDF} Models}. In
  \bibinfo{booktitle}{{\em Proceedings of the Sixteenth Eurographics Conference
  on Rendering Techniques}} {\em (\bibinfo{series}{EGSR '05})}.
  \bibinfo{publisher}{Eurographics Association},
  \bibinfo{address}{Aire-la-Ville, Switzerland, Switzerland},
  \bibinfo{pages}{117--126}.
\newblock
\showISBNx{3-905673-23-1}
\showDOI{%
\url{https://doi.org/10.2312/EGWR/EGSR05/117-126}}


\bibitem[\protect\citeauthoryear{Perlin}{Perlin}{2002}]%
        {Perlin2002}
\bibfield{author}{\bibinfo{person}{Ken Perlin}.}
  \bibinfo{year}{2002}\natexlab{}.
\newblock \showarticletitle{Improving Noise}. In \bibinfo{booktitle}{{\em
  Proceedings of the 29th Annual Conference on Computer Graphics and
  Interactive Techniques}} {\em (\bibinfo{series}{SIGGRAPH '02})}.
  \bibinfo{publisher}{ACM}, \bibinfo{address}{New York, NY, USA},
  \bibinfo{pages}{681--682}.
\newblock
\showISBNx{1-58113-521-1}
\showDOI{%
\url{https://doi.org/10.1145/566570.566636}}


\bibitem[\protect\citeauthoryear{Raymond, Guennebaud, and Barla}{Raymond
  et~al\mbox{.}}{2016}]%
        {Raymond2016}
\bibfield{author}{\bibinfo{person}{Boris Raymond}, \bibinfo{person}{Ga\"{e}l
  Guennebaud}, {and} \bibinfo{person}{Pascal Barla}.}
  \bibinfo{year}{2016}\natexlab{}.
\newblock \showarticletitle{Multi-scale Rendering of Scratched Materials Using
  a Structured {SV-BRDF} Model}.
\newblock \bibinfo{journal}{{\em ACM Trans. Graph.\/}} \bibinfo{volume}{35},
  \bibinfo{number}{4}, Article \bibinfo{articleno}{57} (\bibinfo{date}{July}
  \bibinfo{year}{2016}), \bibinfo{numpages}{11}~pages.
\newblock
\showISSN{0730-0301}
\showDOI{%
\url{https://doi.org/10.1145/2897824.2925945}}


\bibitem[\protect\citeauthoryear{Ritschel, Ihrke, Frisvad, Coppens, Myszkowski,
  and Seidel}{Ritschel et~al\mbox{.}}{2009}]%
        {Ritschel2009}
\bibfield{author}{\bibinfo{person}{Tobias Ritschel}, \bibinfo{person}{Matthias
  Ihrke}, \bibinfo{person}{Jeppe~Revall Frisvad}, \bibinfo{person}{Joris
  Coppens}, \bibinfo{person}{Karol Myszkowski}, {and}
  \bibinfo{person}{Hans-Peter Seidel}.} \bibinfo{year}{2009}\natexlab{}.
\newblock \showarticletitle{{Temporal Glare: Real-Time Dynamic Simulation of
  the Scattering in the Human Eye}}. In \bibinfo{booktitle}{{\em Proceedings
  Eurographics 2009}}.
\newblock


\bibitem[\protect\citeauthoryear{Smits}{Smits}{1998}]%
        {Smits1998}
\bibfield{author}{\bibinfo{person}{Brian Smits}.}
  \bibinfo{year}{1998}\natexlab{}.
\newblock \showarticletitle{Efficiency Issues for Ray Tracing}.
\newblock \bibinfo{journal}{{\em J. Graph. Tools\/}} \bibinfo{volume}{3},
  \bibinfo{number}{2} (\bibinfo{date}{Feb.} \bibinfo{year}{1998}),
  \bibinfo{pages}{1--14}.
\newblock
\showISSN{1086-7651}
\showDOI{%
\url{https://doi.org/10.1080/10867651.1998.10487488}}


\bibitem[\protect\citeauthoryear{Stam}{Stam}{1999}]%
        {Stam1999}
\bibfield{author}{\bibinfo{person}{Jos Stam}.} \bibinfo{year}{1999}\natexlab{}.
\newblock \showarticletitle{Diffraction Shaders}. In \bibinfo{booktitle}{{\em
  Proceedings of the 26th Annual Conference on Computer Graphics and
  Interactive Techniques}} {\em (\bibinfo{series}{SIGGRAPH '99})}.
  \bibinfo{publisher}{ACM Press/Addison-Wesley Publishing Co.},
  \bibinfo{address}{New York, NY, USA}, \bibinfo{pages}{101--110}.
\newblock
\showISBNx{0-201-48560-5}
\showDOI{%
\url{https://doi.org/10.1145/311535.311546}}


\bibitem[\protect\citeauthoryear{Sun, Fracchia, Drew, and Calvert}{Sun
  et~al\mbox{.}}{2000}]%
        {Sun2000}
\bibfield{author}{\bibinfo{person}{Yinlong Sun}, \bibinfo{person}{F.~David
  Fracchia}, \bibinfo{person}{Mark~S. Drew}, {and} \bibinfo{person}{Thomas~W.
  Calvert}.} \bibinfo{year}{2000}\natexlab{}.
\newblock \showarticletitle{Rendering Iridescent Colors of Optical Disks}. In
  \bibinfo{booktitle}{{\em Proceedings of the Eurographics Workshop on
  Rendering Techniques 2000}}. \bibinfo{publisher}{Springer-Verlag},
  \bibinfo{address}{London, UK, UK}, \bibinfo{pages}{341--352}.
\newblock
\showISBNx{3-211-83535-0}
\showURL{%
\url{http://dl.acm.org/citation.cfm?id=647652.732138}}


\bibitem[\protect\citeauthoryear{Trowbridge and Reitz}{Trowbridge and
  Reitz}{1975}]%
        {Trowbridge1975}
\bibfield{author}{\bibinfo{person}{T.~S. Trowbridge} {and}
  \bibinfo{person}{K.~P. Reitz}.} \bibinfo{year}{1975}\natexlab{}.
\newblock \showarticletitle{Average irregularity representation of a rough
  surface for ray reflection}.
\newblock \bibinfo{journal}{{\em J. Opt. Soc. Am.\/}} \bibinfo{volume}{65},
  \bibinfo{number}{5} (\bibinfo{date}{May} \bibinfo{year}{1975}),
  \bibinfo{pages}{531--536}.
\newblock
\showDOI{%
\url{https://doi.org/10.1364/JOSA.65.000531}}


\bibitem[\protect\citeauthoryear{Veach and Guibas}{Veach and Guibas}{1995}]%
        {veach1995optimally}
\bibfield{author}{\bibinfo{person}{Eric Veach} {and}
  \bibinfo{person}{Leonidas~J Guibas}.} \bibinfo{year}{1995}\natexlab{}.
\newblock \showarticletitle{Optimally combining sampling techniques for Monte
  Carlo rendering}. In \bibinfo{booktitle}{{\em Proceedings of the 22nd annual
  conference on Computer graphics and interactive techniques}}. ACM,
  \bibinfo{pages}{419--428}.
\newblock


\bibitem[\protect\citeauthoryear{Wang, Zhao, Tong, Snyder, and Guo}{Wang
  et~al\mbox{.}}{2008}]%
        {Wang2008}
\bibfield{author}{\bibinfo{person}{Jiaping Wang}, \bibinfo{person}{Shuang
  Zhao}, \bibinfo{person}{Xin Tong}, \bibinfo{person}{John Snyder}, {and}
  \bibinfo{person}{Baining Guo}.} \bibinfo{year}{2008}\natexlab{}.
\newblock \showarticletitle{Modeling Anisotropic Surface Reflectance with
  Example-based Microfacet Synthesis}.
\newblock \bibinfo{journal}{{\em ACM Trans. Graph.\/}} \bibinfo{volume}{27},
  \bibinfo{number}{3} (\bibinfo{year}{2008}), \bibinfo{pages}{41:1--41:9}.
\newblock


\bibitem[\protect\citeauthoryear{Yan, Ha\v{s}an, Jakob, Lawrence, Marschner,
  and Ramamoorthi}{Yan et~al\mbox{.}}{2014}]%
        {Yan2014}
\bibfield{author}{\bibinfo{person}{Ling-Qi Yan}, \bibinfo{person}{Milo\v{s}
  Ha\v{s}an}, \bibinfo{person}{Wenzel Jakob}, \bibinfo{person}{Jason Lawrence},
  \bibinfo{person}{Steve Marschner}, {and} \bibinfo{person}{Ravi Ramamoorthi}.}
  \bibinfo{year}{2014}\natexlab{}.
\newblock \showarticletitle{Rendering Glints on High-resolution Normal-mapped
  Specular Surfaces}.
\newblock \bibinfo{journal}{{\em ACM Trans. Graph.\/}} \bibinfo{volume}{33},
  \bibinfo{number}{4}, Article \bibinfo{articleno}{116} (\bibinfo{date}{July}
  \bibinfo{year}{2014}), \bibinfo{numpages}{9}~pages.
\newblock
\showISSN{0730-0301}
\showDOI{%
\url{https://doi.org/10.1145/2601097.2601155}}


\bibitem[\protect\citeauthoryear{Yan, Ha\v{s}an, Marschner, and
  Ramamoorthi}{Yan et~al\mbox{.}}{2016}]%
        {Yan2016}
\bibfield{author}{\bibinfo{person}{Ling-Qi Yan}, \bibinfo{person}{Milo\v{s}
  Ha\v{s}an}, \bibinfo{person}{Steve Marschner}, {and} \bibinfo{person}{Ravi
  Ramamoorthi}.} \bibinfo{year}{2016}\natexlab{}.
\newblock \showarticletitle{Position-normal Distributions for Efficient
  Rendering of Specular Microstructure}.
\newblock \bibinfo{journal}{{\em ACM Trans. Graph.\/}} \bibinfo{volume}{35},
  \bibinfo{number}{4}, Article \bibinfo{articleno}{56} (\bibinfo{date}{July}
  \bibinfo{year}{2016}), \bibinfo{numpages}{9}~pages.
\newblock
\showISSN{0730-0301}
\showDOI{%
\url{https://doi.org/10.1145/2897824.2925915}}


\end{thebibliography}
\begin{appendix}
\section{Gaussian weighted spatial phases}\label{app:spatial-phases}
{
The integration of the spatial phases of the scratches (and masks) relies on the following assumptions: First, the width of a scratch is negligible compared to its length 
with respect to the coherence area. Second, the scratch segments we integrate are lines. Third, the profile $\mathcal{P}$ does not change over a segment. 
To regain spatial resolution we apply a spatial filter kernel on the surface which weights each phase $\Phi^{(k)}(\tancoord)$ (see \Sec~\ref{sec:theory-scratch-brdf}) according to its position relative to the 
point of intersection (the origin of the footprint, see  \Fig~\ref{fig:shading-geometry}) and approximates the coherence function. This provides us with a closed-form solution of the integral of \Eq~\ref{eq:single-scratch-otf-fourier-transform} 
given by
\begin{equation}
\eta_k = \!\int\!\!d\tancoord\,\mathcal{G}(\tancoord)\,\Phi(\tancoord)\\ \nonumber 
\end{equation}
We represent our scratches as the relative position $\cvec{r}'(t) = \cvec{r}'_0 + \tancoord \cdot \uvec{t}' - \cvec{x}'_0;\,\tancoord \in [-L/2, L/2]$ and $\cvec{r}_0'(t) = \cvec{r}'_0 - \cvec{x}'_0$ where $L$ is the length of the scratch and 
the prime denotes the coordinates in tangent space. The integral to solve then reads
\begin{eqnarray}
\label{eq:spatial-phases-result}
\mkern-36mu&=&\int\!\!d\tancoord\, \Phi^{(k)}(\tancoord)\,e^{-|\cvec{r}'(\tancoord)|^2/(2\sigma^2)}\\ \nonumber
\mkern-36mu&=&\int\!\!d\tancoord\, e^{-2 \pi i (\cvec{r}'(\tancoord))\cdot\cvec{\xi}'}\,e^{-|\cvec{r}'(\tancoord)|^2/(2\sigma^2)}\\ \nonumber
\mkern-36mu&=& e^{-2 \pi i \cvec{r}'_0(\tancoord)\cdot\cvec{\xi}'}\,e^{-\frac{|\cvec{r}'_0|^2}{2\sigma^2}}\int\!\!d\tancoord\, e^{-2 \pi i \tancoord \cvec{\xi}_1'}\,e^{-\frac{\tancoord^2 + 2\tancoord(\uvec{\tancoord}' \cdot \cvec{r}_{r,0}')}{2\sigma^2}}\\ \nonumber
\mkern-36mu&=& c_0 \cdot \left[ \mathrm{erf}\left(\frac{a_0 + L/(2\sigma)}{\sqrt{2}}\right) - \mathrm{erf}\left(\frac{a_0 - L/(2\sigma)}{\sqrt{2}}\right)\right]\nonumber
\end{eqnarray}
where $L$ is the total length of the scratch and
\begin{eqnarray}
  a_0 &=& 2\pi i \sigma \cvec{\xi}'_1 + \frac{(\uvec{\tancoord}' \cdot \cvec{r}'_0)}{\sigma}\\ \nonumber 
  c_0 &=& \sqrt{\pi/2}\,\sigma\,e^{e + i f}\\ \nonumber
  e &=& -\frac{|\cvec{r}'_0|^2}{2\sigma^2} - 2\pi^2\sigma^2\cvec{\xi}_1'^2 + \frac{(\uvec{\tancoord}' \cdot \cvec{r}'_0)^2}{2\sigma^2}\\ \nonumber
  f &=& 2\pi i \cvec{\xi}'_1 (\uvec{\tancoord}' \cdot \cvec{r}'_0) - 2\pi i (\cvec{r}'_0\cdot\cvec{\xi}')
\end{eqnarray}
}
{
\section{Scratch and mask profiles}\label{app:profiles}
We separate the spatial and the spectral component of the scratches which enables us to drive the reflection properties of a scratch by varying its 1d transfer function via the profile $\mathcal{P}_\scratch(b)$. 
As we need to compute the Fourier transform of $\mathcal{T}_\scratch(b)$ it is convenient to choose profile functions that lead to a closed-form solution and are drivable by the geometric parameters 
 width $W$ and depth $D$. We note that it would in general be possible to replace this function by a lookup table. The transfer function for a specific profile in general reads:
\begin{equation}
 \mathcal{T}(b) = A(b)\,e^{-4 \pi i \mathcal{P}(b)/\lambda}
\end{equation}
where $A(b)$ contains the Fresnel term and $b$ is the bitangential coordinate in the ($\uvec{b}$,$\uvec{n}$)-plane (c.f. fig~\ref{fig:shading-geometry}).

The mask transfer function $\mathcal{T}_\mask(b)$ then is defined by  $A_\mask(b) = A_\base$ without any phase changes because 
no height deviations are present. This yields the Fourier transform of a rectangular function
\begin{eqnarray}
 \mathcal{F}\left\{ \mathcal{T}_\mask(b)\right\}_{\cvec{\xi}'_2} &=& \mathcal{F}\left\{ A_\base\,\mathrm{rect}\left(\frac{b}{W}\right)\right\}_{\cvec{\xi}'_2}\\ \nonumber
 &=& A_\base\,W \cdot \mathrm{sinc}\left(\pi W\cvec{\xi}'_2 \right)
\end{eqnarray}
which is always the case for the mask profile. On the other hand we are able to choose an arbitrary scratch profile. For simplicity we concentrate on two different 
profile functions which are driven by the scratches' widths and depths. The simplest case is a rectangular profile with constant depth, i.e.,
\begin{eqnarray}
\mkern-36mu \mathcal{F}\left\{ \mathcal{T}_\scratch^\mathrm{rect}(b)\right\}_{\cvec{\xi}'_2} &=& \mathcal{F}\left\{A_\scratch\,\mathrm{rect}\left(\frac{b}{W}\right) \right\}_{\cvec{\xi}'_2}\Phi_D \\ \nonumber
\mkern-36mu &=& A_\scratch\,W \cdot \mathrm{sinc}\left(\pi W \cvec{\xi}'_2 \right) \Phi_D
\end{eqnarray}
and constant depth-phase 
\begin{equation}
 \Phi_D = e^{-4\pi i D/\lambda}
\end{equation}
For triangular profiles, we obtain
\begin{align}
\begin{split}
 &\mathcal{F}\left\{ \mathcal{T}_\scratch^\mathrm{tri}\right\}_{\cvec{\xi}'_2} \\
 =& \mathcal{F}\left\{ A_\scratch\,\mathrm{rect}\left(\frac{b}{W}\right) \cdot e^{4\pi i D/\lambda\,(1 - |\frac{b}{W/2}|)}\right\}_{\cvec{\xi}'_2} \\
 =& B \cdot \left( 1 - e^{\pi i ({W\cvec{\xi}'_2} - \frac{4D}{\lambda})}\right)\\ 
 &+ C \cdot \left(e^{-\pi i ({W\cvec{\xi}'_2} + \frac{4D}{\lambda})} - 1\right)\\
 \end{split}
\end{align}
with 
\begin{eqnarray}
 B &=& A_\scratch\,\frac{i\,\Phi_D}{2\pi ({\cvec{\xi}'_2} - \frac{4D}{W\lambda})}\\
 C &=& A_\scratch\,\frac{i\,\Phi_D}{2\pi ({\cvec{\xi}'_2} + \frac{4D}{W\lambda})}
\end{eqnarray}
}
\end{appendix}

\end{document}